\documentclass[3p,authoryear]{elsarticle}

\journal{Icarus}

\usepackage{natbib}

\usepackage{graphicx}

\usepackage{amssymb}
\usepackage{amsmath}
\usepackage{color}
\usepackage[usenames,dvipsnames]{xcolor}
\usepackage{ulem}
\usepackage{doi}
\usepackage{hyperref}
\hypersetup{colorlinks=true, urlcolor=blue}

\usepackage{lineno}

\mathchardef\mhyphen="2D

\begin{document}

\begin{frontmatter}



\title{In-flight Calibration of the Dawn Framing Camera\tnoteref{label1}\tnoteref{label2}}
\tnotetext[label1]{\doi{10.1016/j.icarus.2013.07.036}}
\tnotetext[label2]{\copyright 2017. This manuscript version is made available under the CC-BY-NC-ND 4.0 licence:\\ \url{https://creativecommons.org/licenses/by-nc-nd/4.0/}}


\author[DLR,MPS]{S.E.~Schr\"oder}
\author[MPS]{T.~Maue}
\author[MPS]{P.~Guti{\'e}rrez Marqu{\'e}s}
\author[DLR]{S.~Mottola}
\author[UCLA]{K.M.~Aye}
\author[MPS]{H.~Sierks}
\author[IGEP]{H.U.~Keller}
\author[MPS]{A.~Nathues}

\address[DLR]{Deutsches Zentrum f\"ur Luft- und Raumfahrt (DLR), 12489 Berlin, Germany}
\address[MPS]{Max-Planck-Institut f\"ur Sonnensystemforschung (MPS), 37191 Katlenburg-Lindau, Germany}
\address[UCLA]{Department for Earth and Space Sciences (ESS), University of California, Los Angeles, CA 90095-1567, U.S.A.}
\address[IGEP]{Institut f\"ur Geophysik und Extraterrestrische Physik (IGEP), Technische Universit\"at Braunschweig, 38106 Braunschweig, Germany}

\begin{abstract}

We present a method for calibrating images acquired by the Dawn Framing Camera (FC) that is based on the results of an in-flight calibration campaign performed during the cruise from Earth to Vesta. We describe this campaign and the data analysis in full. Both the primary camera FC2 and the backup camera FC1 are radiometrically and geometrically calibrated through observations of standard stars, star fields, and solar system objects. The calibration in each spectral filter is accurate to within a few percent for point sources. Geometric distortion, small by design, is characterized with high accuracy. Dark current, monitored on a regular basis, is very low at flight operational temperatures. Out-of-field stray light was characterized using the Sun as a stray light source. In-field stray light is confirmed in narrow-band filter images of Vesta. Its magnitude and distribution are scene-dependent, and expected to contribute significantly to images of extended objects. Description of a method for in-field stray light correction is deferred to a follow-up paper, as is a discussion of the closely related topic of flat-fielding.

\end{abstract}


\end{frontmatter}



\section{Introduction}
\label{sec:introduction}

The Dawn spacecraft has finished its year-long mission around main-belt asteroid 4~Vesta and is on its way to its next target Ceres \citep{R07,R12}. The on-board Framing Camera (FC) acquired tens of thousands of images of the Vesta surface. In this paper we present a method for calibrating these images that is based on an analysis of in-flight data, acquired during the cruise to Vesta. This in-flight calibration builds on the pre-launch laboratory calibration reported by \citet{S11}. The FC is not only a scientific instrument, but also serves optical navigation purposes. Being critical to the success of the mission, redundancy was judged to be essential. Thus, Dawn has two cameras: primary camera FC2 (formerly flight model \#2) and backup camera FC1 (formerly flight spare \#1). We discuss the radiometric and geometric calibration of both instruments. We also describe the format of different types of images acquired by the FC, and the details of various in-flight calibration campaigns performed up to Vesta arrival. This paper aims to allow the reader to calibrate the raw images that are archived in the NASA Planetary Data System\footnote{\url{http://pds.nasa.gov}} (PDS). Master dark frames, also archived in the PDS, can be used for this purpose.

The Framing Cameras are mounted on the instrument deck of the spacecraft \citep{S11}. They are equipped with one clear and seven narrow band filters (full width at half maximum is typically 40~nm). The following aspects of the camera design must be considered when calibrating the images. The FC has a protective door, placed on top of the baffle, which is opened at the start of an imaging campaign and closed afterwards. The detector is a front-illuminated, frame transfer CCD. The images are digitized to 14-bit depth, so that the full range is 0 to 16383 data numbers (DN). The system gain in the FC was adjusted such that the analog-to-digital saturation approximately corresponds to the pixel full well. Image exposures are acquired using an electronic shutter, eliminating the need for a, potentially unreliable, mechanical shutter. Electronic shuttering makes use of the anti-blooming architecture of the CCD, which differs from the method of fast-shifting the accumulated charge before the exposure that is commonly used for frame-transfer CCDs. When the door is open, light falls on the CCD. When not exposing, the CCD is in ``idle mode'', in which the parallel and serial registers are continuously clocked to clear the accumulated charge. At the start of an exposure, idle mode is terminated, and the charge still present in the parallel register is cleared by antiblooming drains. After the exposure has finished, the image is transferred from the CCD active area into a storage area for read-out. The CCD continues to be exposed during this transfer, creating a smear that must be removed in the calibration process. The anti-blooming gates reduce the light-sensitive area of the pixel, known as the fill factor. Consequently, repeated observations of a point source, such as a photometric standard star, may result in very different measured intensities, depending on the sub-pixel position of the centroid of the source point spread function (PSF). This significantly complicates the derivation of an accurate radiometric calibration. Hence a large part of this paper is devoted to this subject. FC images are routinely compressed on board, as documented in the image header \citep{S11}. Compression of images acquired for calibration purposes is generally lossless. Science images are compressed lossy, but often with a compression ratio that approximates lossless quality. When the ratio is set sufficiently low ($\sim1.8$ for well illuminated images of the Vesta surface), the compression artifacts are not larger than 1~DN. At higher ratios, compression effectively smoothes photon noise.

This paper builds on the results of the ground calibration. Several aspects were not reassessed during flight, and are summarized here. Details of the ground calibration can be found in \citet{S11}, except where indicated. At the typical flight operational CCD temperature (220~K), the gain is $17.7 \pm 0.3$ electrons per DN \citep{C06}. Hence, the expected photon noise in an uniformly illuminated area in an image with average (clean) signal level $\bar{s}$ (in DN) is $\sigma_s = (\bar{s} / 17.7)^{0.5}$~DN, resulting in a signal-to-noise (S/N) of $(17.7 \bar{s})^{0.5}$. Deviation from linearity, which includes contributions by the CCD, analog processor, and analog-to-digital converter, is smaller than 1\% for signal levels below 12~kDN \citep{C06,Sc11}. As the vast majority of the images was exposed well below this level, we ignore non-linearity in our analysis. A digitization error of 0.5~DN is part of the $1.14 \pm 0.05$~DN read-out noise \citep{C06}, which is monitored in-flight (Sec.~\ref{sec:bias}). \citet{C06} did not find evidence for coherent noise. The PSF diameter for filters F1 through F8 was calculated to be 1.4, 0.9, 1.2, 1.0, 1.2, 0.9, 1.1, and 2.2 pixels in the image center (80\% encircled energy). An analysis of images of the inside of an integrating sphere -- a common technique to produce flat fields -- suggests that FC narrow band images are significantly affected by in-field stray light. For this reason, the lab images as recorded cannot be used as flat fields. We attempt to verify the occurrence of this phenomenon in-flight, both in quantitative and qualitative sense, and discuss the implications for the calibration (Sec.~\ref{sec:in-field}). A method for stray light correction and flat field revision is the subject of a follow-up paper, which will also include a list of bad pixels.

One aspect of the in-flight calibration campaign, optical alignment, is not addressed in this paper, as it is not part of the image calibration process. The camera boresight pointing was derived with high accuracy from in-flight data (star field images). The alignment of the FC and other Dawn instruments is documented in the relevant {\sc spice} frame kernels\footnote{See \url{http://naif.jpl.nasa.gov/naif/} for more information on SPICE. The latest FC frame kernel at the time of writing is \texttt{dawn\_v11.tf}.}.

\section{Framing Camera imagery}

For diagnostic purposes, the FC CCD has regions around the active area that are covered by opaque material. The FC is able to return images in different formats, related to how these regions are treated. It can also acquire special types of images for calibration purposes, like dark exposures and images of the inside of the camera door illuminated by a calibration lamp. Here we describe the different image formats and types. Note that each image has a unique identifying number, which, in this paper, is printed in {\bf bold} font. An image is taken through one of eight filters, identified as F1 (clear filter), and F2 through F8 (narrow-band filters).

\subsection{Image formats}
\label{sec:image_formats}

\begin{figure}
\centering
\includegraphics[width=\textwidth,angle=0]{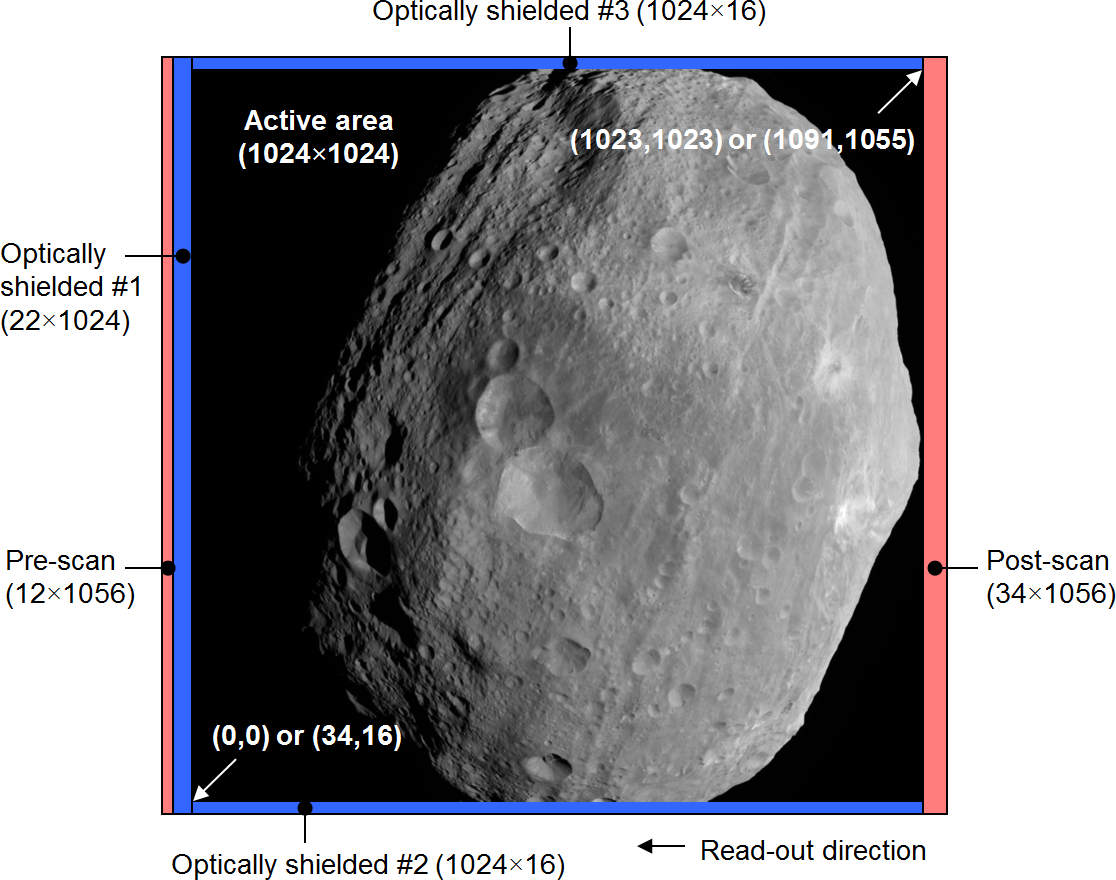}
\caption{Layout of a {\bf full-full frame} FC image. Shown are the active area, in this case containing clear filter image {\bf 3283}, the three optically shielded regions (dark blue), and the pre- and post-scan regions (red). Area size is indicated in (columns $\times$ rows). The coordinates of the pixels in the lower left and upper right corner of the active area are (0,0) and (1023,1023) in a {\bf full frame} {\sc image} object, and (34,16) and (1057,1039) in a {\bf full-full frame} {\sc image} object. The storage area is located below the area shown here. The horizontal (read-out) direction is referred to as the sample- or $x$-direction, the vertical direction is the line- or $y$-direction.}
\label{fig:image_frames}
\end{figure}

The native format for FC data files complies with the PDS standard, with one or more image objects embedded in the file and an attached header describing the contents. The raw and decompressed images are referred to as {\bf level~1a} data products \citep{S11}. Level~1a image objects are stored as 16-bit integers LSB, as described in the header. An FC data file always contains an image object called {\sc image}, to allow PDS-compatible applications to display the images correctly. The size and format of any additional image objects is determined by the image acquisition command and the associated cropping options. Only two cropping options are associated with regular FC observations. The standard {\bf full frame} contains an {\sc image} object of $1024\times1024$ pixels representing the illuminated part of the CCD, called the active area, and four additional objects of different dimensions representing the bias and optical shield regions (Fig.~\ref{fig:image_frames}). The shielded regions can be used to monitor the dark current. However, we prefer to use a dark current model because the shielded regions have a dark current floor that is slightly different from that of the active area \citep{C06}. The naming of these objects adhered to the following convention during the Vesta campaign: Objects {\sc frame\_2\_image}, {\sc frame\_3\_image}, {\sc frame\_4\_image}, and {\sc frame\_5\_image} represent the pre-scan, shielded \#1, shielded \#2, and shielded \#3 regions, respectively (note: the dimensions of these objects are not identical to those of the regions). The {\bf full-full frame} contains one object of $1088\times1056$ pixels ({\sc frame\_0\_image}), another of $4\times1056$ pixels ({\sc frame\_6\_image}), and an {\sc image} object of $1092\times1056$ pixels, created from the first two objects during ground processing. The location of each object within the full logical CCD area of $1092\times1056$ pixels is included in the attached header.

Occasionally, images are acquired in {\bf windowed} mode, for example when observing standard stars. The level~1a image then contains an {\sc image} object of $256\times256$ pixels representing the window, and four additional objects associated with the aforementioned bias and optical shield regions. The location of the window within the active area can be retrieved from the {\sc first\_line} and {\sc first\_line\_sample} keywords in the object header.

\subsection{Diagnostic image types}

In addition to the regular science image acquisitions, there are a number of image types that the camera can produce for the purpose of diagnosing its health status and measuring its performance. These images are identified in the attached PDS header of each file by the keyword {\sc dawn:image\_acquire\_mode} having a value other than {\sc normal}. They serve calibration purposes and are not calibrated themselves, i.e.\ exist only as level~1a products. We distinguish the following types:

{\bf Serial readout.} In this mode, the CCD serial register is read through the output amplifier without shifting the charge from the active area into the storage area. These images are identified with the keyword value {\sc serial} and provide full-full frames.

{\bf Storage readout.} This diagnostic mode measures the dark current generated in the storage area of the CCD. As with the serial readout, the charge created in the exposed (active) area is not transferred to the storage area. At the start of the exposure the storage area is cleared of charge. At the end of the exposure the amplifier reads out the charge accumulated in the storage area for the duration of the exposure. These images are identified as {\sc storage} and provide full-full frames as well.

{\bf Dark.} A dark frame is an image acquired with the front door closed. The purpose of this measurement is determining the dark current generation rate in each pixel of the CCD active area. Given that this rate is very small for most of the pixels in the in-flight operational temperature range, dark frames are usually taken with long exposure times of tens or even hundreds of seconds. These images are identified as {\sc dark} and provide standard full frames. Note that long dark exposures are strongly affected by cosmic ray hits.

{\bf Calibration lamp.} LEDs mounted in the FC front baffle permit the controlled illumination of the CCD when the door is closed \citep{S11}. These lamps are subject to variations in their output depending on the temperature, the phase of the mission (age of the lamps), and the voltage in the power rail. However, even a varying light source is enough to determine variations in responsivity between neighboring pixels. The FC1 and FC2 calibration lamp brightness distributions are very different. While these images are most certainly not flat fields, they are identified as {\sc flatfield} and provide standard full frames.

\section{In-flight calibration campaigns}

This section presents an overview of the observational campaigns in which the data were acquired that we analyze in this paper. These campaigns were conducted with the exclusive purpose of instrument calibration. For clarity, campaign names are displayed in {\it italics}. For historical reasons we mention that the {\it Mars Gravity Assist} in February 2009 presented an important opportunity to improve the FC radiometric calibration, as it was the first time that the camera would have an extended object in its field-of-view after launch. Unfortunately, due to a spacecraft safing near closest approach \citep{RM10} only a handful of FC2 images were transmitted, most of which feature the unilluminated side of the planet. Only five images were acquired on which surface details can be distinguished, all acquired in the late local evening.

\subsection{Initial Checkout Operations}
\label{sec:ICO}

Performed right after launch, the Initial Checkout Operations ({\it ICO}) of the Framing Cameras consisted of three individual campaigns, called {\it Functional}, {\it Performance}, and {\it Calibration}. During the {\it Functional} campaign the operability of the FC was demonstrated by acquiring a set of diagnostic images with closed front cover, as well as by observing a random patch of sky in each filter with three different exposure times (1/30/300~s). During {\it Performance} the camera was pointed at two photometric standard stars, 20~Cep and Vega, to verify the radiometric calibration that was based on lab measurements. The stars were observed in all filters with two different exposure times per filter. Vega was also imaged while the spacecraft performed a slow slew in order to mitigate the effects of the reduced pixel fill factor \citep{S11}. Repeated short observations of a point source can yield intensities that differ by a factor of two, depending on the location of the centroid of the PSF within the pixel (discussed in more detail in Sec.~\ref{sec:standard_stars}). Slewing will help to sample different parts of the pixel. For {\it Calibration}, the target selection was different for FC1 and FC2 due to flight rule constraints. FC1 observed the following targets in all filters: 73~Cet, 51~Peg, and NGC 3532. The latter is an open cluster in the constellation Carina, known as the Wishing Well Cluster. The target selection for FC2 was: 42~Peg, 51~Peg, NGC~3532, Saturn, and the constellation Cassiopeia. Details of the {\it ICO} observations are listed in Table~\ref{tab:ICO}.

\subsection{Semi-annual checkouts}
\label{sec:semi-annuals}

Camera checkouts are performed twice per year during cruise. Their purpose is threefold: verifying the integrity of the mechanical and electronic components, exercising the camera mechanisms, and monitoring the camera performance \citep{S10}. The latter refers to monitoring the development of hot and blind pixels, the bulk dark current, the residual charge (explained in Sec.~\ref{sec:extra_charge}), and the radiometric and geometric properties. We established that blind pixels are most easily identified in averaged images of Vesta rather than images of the calibration lamp, and a discussion of this topic will be included in a follow-up paper. The checkout comes in two flavors: with and without dedicated pointing. The {\it pointed checkout} has the camera pointing to various celestial objects, such as photometric standard stars and star fields, for calibration purposes. A prime and backup target are selected for each of the following targets: standard star, solar analog star, mission target, and star field (Table~\ref{tab:semi-annual}). If the prime target cannot be observed at the time of the checkout due to attitude constraints then the backup target is observed. The standard star is imaged while the spacecraft is slewing in order to improve the photometric accuracy of the measurements. A calibration target is observed through all filters, the photometric standard star at nine different positions around the center of the CCD, the other targets twice at the same position. During the {\it non-pointed checkout} the camera observes only random patches of sky, at whatever the spacecraft happens to be pointed at the time of acquisition. A few checkout images are lossy compressed with different compression algorithms and ratios to monitor the flight software performance. Some star field images were found to be overexposed at the location of bright stars, with associated charge bleeding on the CCD. An analysis of these images is not included in this paper, as only a handful of pixels was overexposed during the Vesta campaign.

Each checkout sequence is executed once a year for FC2, whereas for FC1 the non-pointed checkout is executed twice a year. The diagnostic images have the full-full frame format, i.e.\ are sized $1092\times1056$ pixels. The standard star and solar analog images are returned in the windowed format sized $256\times256$ pixels. Data of semi-annual checkout {\it DC041} are included in the analysis in Sec.~\ref{sec:geom_dist}.

\subsection{Stray light campaign}
\label{sec:stray_light_campaign}

On 31 March 2009 FC2 performed observations dedicated to characterizing out-of-field stray light. It acquired a series of 1/10/100~s exposures in all filters with the door open and the Sun at off-axis angles ranging from $90.0^\circ$ to $30.0^\circ$ with $2.5^\circ$ decrements. Flight rules restricted the minimum off-axis Sun angle to $30^\circ$. The results of this campaign are detailed in Sec.~\ref{sec:out-of-field}.

\section{Image calibration steps}

In this section we detail how to convert the raw level~1a image in data number (DN) units, to a radiometrically calibrated image in physical (SI) units. Calibrating FC images involves the usual steps of bias and dark current subtraction, and division by a flat field. In some cases additional processing steps are required.

\subsection{Overview}

The raw image $\mathbf{W}^i$ recorded by the CCD (in DN) for filter $i$ ($i=1$-$8$) and an exposure time $t_{\rm exp}$ (in seconds) is a sum of several terms, each of which is discussed in a separate section below:
\begin{equation}
\mathbf{W}^i(t_{\rm exp}) = t_{\rm exp} [\mathbf{C}^i + \mathbf{D} + \mathbf{I}^i(\mathbf{C}) + \mathbf{O}^i] + \mathbf{X}(\mathbf{C}) + \mathbf{S}(\mathbf{C}) + b.
\label{eq:image}
\end{equation}
The first four terms on the right side of the equation are in units of DN~s$^{-1}$. $\mathbf{C}^i$ is the ``clean'' image, i.e.\ the charge rate due to the observed scene. $\mathbf{D}$ is the dark current image. $\mathbf{I}^i$ is the in-field stray light contribution for filter $i$. In-field stray light is considerable for the narrow-band filters, where it depends on the observed scene, but negligible in the clear filter; hence we assume $\mathbf{I}^1 = \mathbf{0}$. $\mathbf{O}^i$ is the out-of-field stray light contribution for filter $i$; it depends of the scene outside the field-of-view. The stray light terms are explicitly labeled with filter number $i$ because the patterns are different for each filter. $\mathbf{X}$ is the residual charge, which has not been observed in FC2 images, but is present in FC1 images, where it is a function of the clean charge rate. $\mathbf{S}$ is the smear contribution due to the time it takes to transfer the image from the active area of the CCD into the storage area; it is also a function of the clean charge rate. Finally, $b$ is the bias, a single number for the entire frame. The purpose of the calibration is to reconstruct the clean image $\mathbf{C}^i$ from the raw image $\mathbf{W}^i$, and convert its units of DN~s$^{-1}$ to physical units of radiance, resulting in the radiometrically calibrated image $\mathbf{L}^i$.

\subsection{Bias}
\label{sec:bias}

The readout electronics adds a voltage bias to the CCD video output to prevent the occurrence of negative signals due to noise. This effectively adds a certain, ideally identical, value to all pixels in the image that is called the bias ($b$ in Eq.~\ref{eq:image}). This bias has to be subtracted from the raw image as a first step in the calibration. The leftmost 12 columns of the full-full image frame ($1092\times1056$ pixels), the {\it pre-scan} region, are included as the {\sc frame\_2\_image} object in a full frame FC image. They do not represent a physical area on the CCD, but instead are filled with values from leading pixels in the serial register. These values are in floating point format, although early in the mission they were returned as integers. The average of the pre-scan region provides a good estimate of the bias level, while the standard deviation is a measure of the readout noise, confirmed to be $1.14 \pm 0.05$~DN. The typical FC bias is between 250 and 300 DN, with the FC1 bias around 10~DN higher than that of FC2. Zero-second exposures with the camera cover closed are routinely acquired during flight. Apart from the occasional presence of cosmic rays they show no variability over the frame. On average, the data numbers are slightly higher in the active area than in the bias region (leftmost 12 columns), consistent with accumulation of dark current in the storage area during read-out. This justifies our choice for a single bias value for the entire frame.

\subsection{Dark current}
\label{sec:dark}

We distinguish between the dark current of the bulk of the CCD pixels (``dark current floor'') and that of ``hot pixels'', which is substantially higher. The dark current image $\mathbf{D}$ at CCD temperature $T_{\rm CCD}$ (in K) is calculated from a master dark image $\mathbf{M}$ and the dark current floor $B$ as
\begin{equation}
\mathbf{D}(T_{\rm CCD}) = \frac{B(T_{\rm CCD})}{B(T_{\rm ref})} \mathbf{M}(T_{\rm ref}),
\label{eq:ref_dark}
\end{equation}
in which $T_{\rm ref}$ (in K) is the reference temperature of the master dark frame. All dark currents in Eq.~\ref{eq:ref_dark} are in units of DN~s$^{-1}$. The dark current floor can be modeled with the Arrhenius law \citep{W02}:
\begin{equation}
B(T) = a e^{-b / (k_{\rm B} T)},
\label{eq:bulk_dark}
\end{equation}
where $a$ and $b$ are constants and $k_{\rm B} = 1.38065 \times 10^{-23}$ m$^2$ kg s$^{-2}$ K$^{-1}$ is the Boltzmann constant. As this simple model fits laboratory measurements over a wide range of temperatures \citep{S11}, a more complicated expression is not required. The change of dark current with temperature is determined by the $b$-constant, which was found to be $1.018\times10^{-19}$ m$^2$ kg s$^{-2}$ from pre-launch measurements. We track the dark current floor using the average of row 1000 of a dark image, excluding cosmic rays and hot pixels. It has been monitored since launch. The $a$-``constant'' jumped to a higher value right after launch, and has increased since (Fig.~\ref{fig:bulk_dark}). At the time of writing, the values are $a = 1.64\times10^{13}$ and $2.46\times10^{13}$ DN~s$^{-1}$ for FC1 and FC2, respectively. The master dark frames are constructed as the median of a sequence of dark exposures (typically $n = 9$). Before applying the median filter, the individual dark frames, which are heavily affected by cosmic ray hits, are scaled to the same reference temperature by means of Eq.~\ref{eq:bulk_dark}. The reference temperature reflects the typical operational temperature of the FC1 and FC2 CCD (222~K and 219~K, respectively).

The master dark frames, archived in the PDS, reveal the evolution of the number of hot pixels. We define a hot pixel as one that has a dark current in excess of the average by $5\sigma$, $\sigma$ being the standard deviation of the dark current floor excluding hot pixels (calculated in iterative fashion). We compare the master dark frames of the {\it ICO Performance \& Calibration} (February 2008) and {\it VTH} (September 2011) campaigns for FC1, and those of the {\it ICO Calibration} campaign (December 2007) and the {\it Low Altitude Mapping Orbit} ({\it LAMO}, orbit C13, March 2012) for FC2. As the dark current floor slowly increases with time, the hot pixel definition limit has also increased. It has gone from 0.088 to 0.11~DN~s$^{-1}$ for FC1 in 43 months, and from 0.080 to 0.10~DN~s$^{-1}$ for FC2 in 51 months. The percentage of hot pixels has increased from 0.4\% to 2.5\% for FC1, and from 0.2\% to 2.4\% for FC2. That said, the number of pixels that have a dark current over 1~DN~s$^{-1}$ remains small, only 812 for FC1 (0.08\%) and 537 for FC2 (0.05\%) for the last of the two campaign investigated.

While the dark current has essentially doubled since launch, both in terms of the dark current floor and the number of hot pixels, it remains low at its current level of 0.05-0.06~DN~s$^{-1}$ for the typical in-flight operational CCD temperature of each camera. As typical images of the asteroid surface have exposure times measured in tens to hundreds of milliseconds and signal levels of several thousands of DNs, the dark current contribution can generally be ignored, except if one is interested in shadowed crater floors or when searching for moons in long exposed images around the asteroid.

\subsection{Residual charge}
\label{sec:extra_charge}

During the Vesta campaign several images of the asteroid have been obtained with the FC1 camera for diagnostic purposes. FC1 is very similar to FC2 with one important exception: FC1 images are affected by residual charge. This charge is mostly apparent in acquisitions of extended sources, but it affects point sources as well. The amount varies from pixel to pixel, as it depends both on the input flux and on the individual characteristics of each pixel. Residual charge is the result of an imperfect clearing of the CCD before the start of the exposure, which is controlled by the anti-blooming gates. As explained in the introduction, immediately before the start of an image exposure the charge in the pixel wells is dumped to an anti-blooming drain. The efficiency with which this charge is removed is controlled by the voltage applied to the anti-blooming gates. It appears that this voltage was not set to the optimal value for FC1, leading to incomplete removal. Therefore, FC1 images of Vesta (or Ceres) need to undergo an additional calibration step: subtraction of the residual charge image $\mathbf{X}$ from the raw image. To date, no trace of residual charge has been found in images of the mission's primary camera, FC2.

\begin{figure}
\centering
\includegraphics[width=8cm,angle=0]{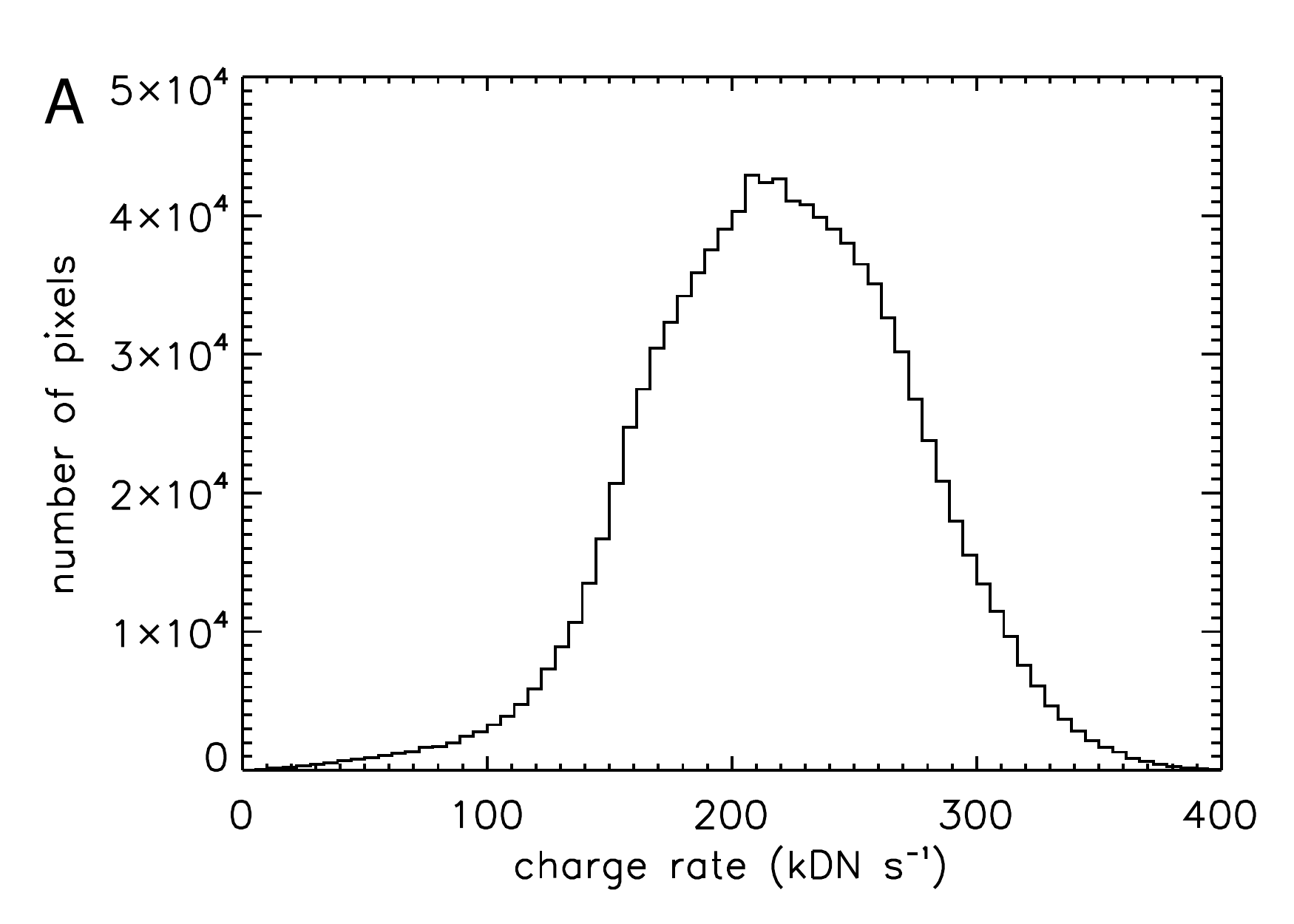}
\includegraphics[width=8cm,angle=0]{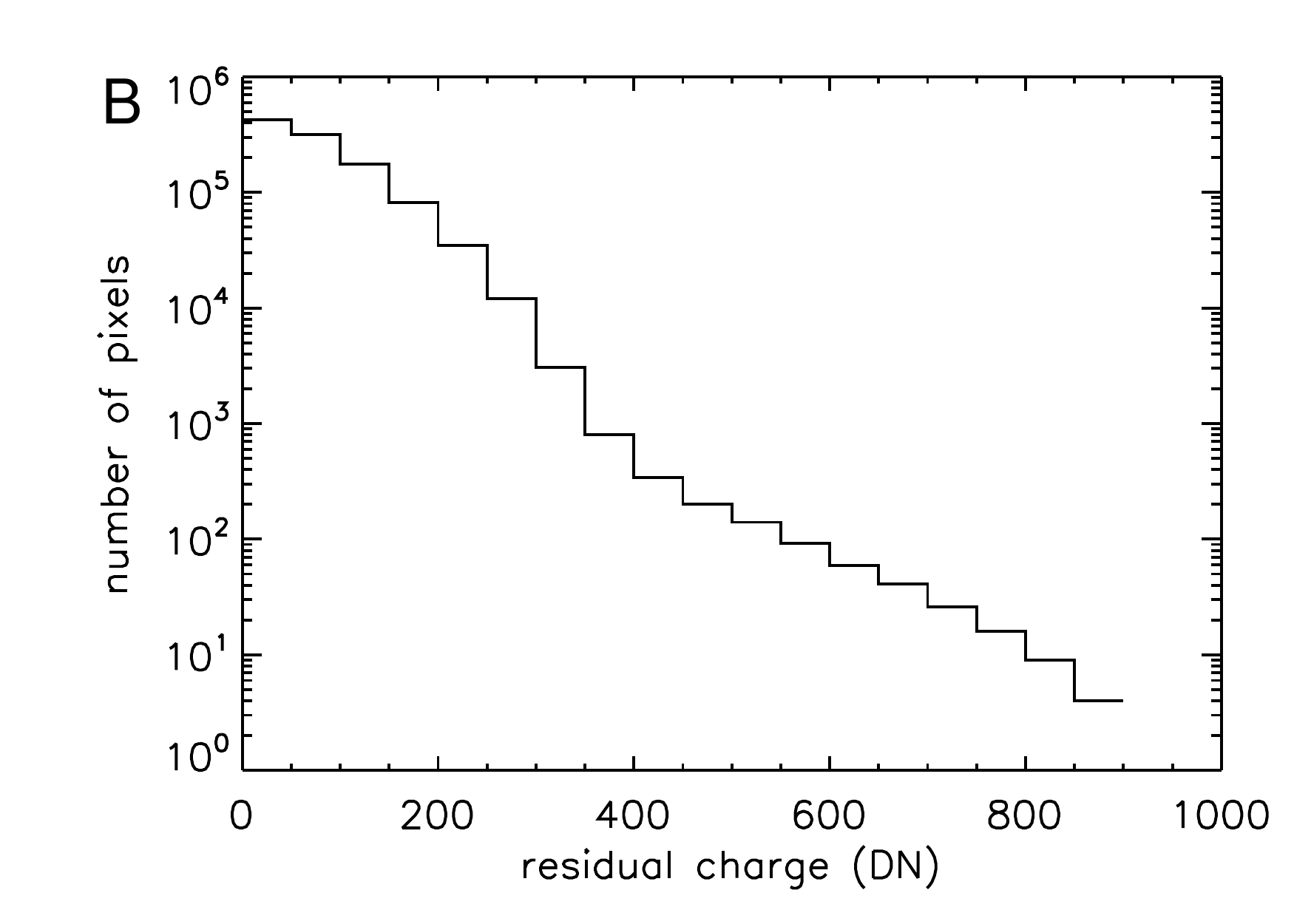}
\caption{Residual charge in FC1 image {\bf 1241}, with an exposure time of 0.009~s. The surface of Vesta completely fills the FOV. The median charge of the raw image is 2438~DN. {\bf A}:~Histogram of the charge rate of the (clean) image. {\bf B}:~Histogram of the residual charge present on the CCD at the start of the exposure.}
\label{fig:extra_charge}
\end{figure}

\begin{figure}
\centering
\includegraphics[width=8cm,angle=0]{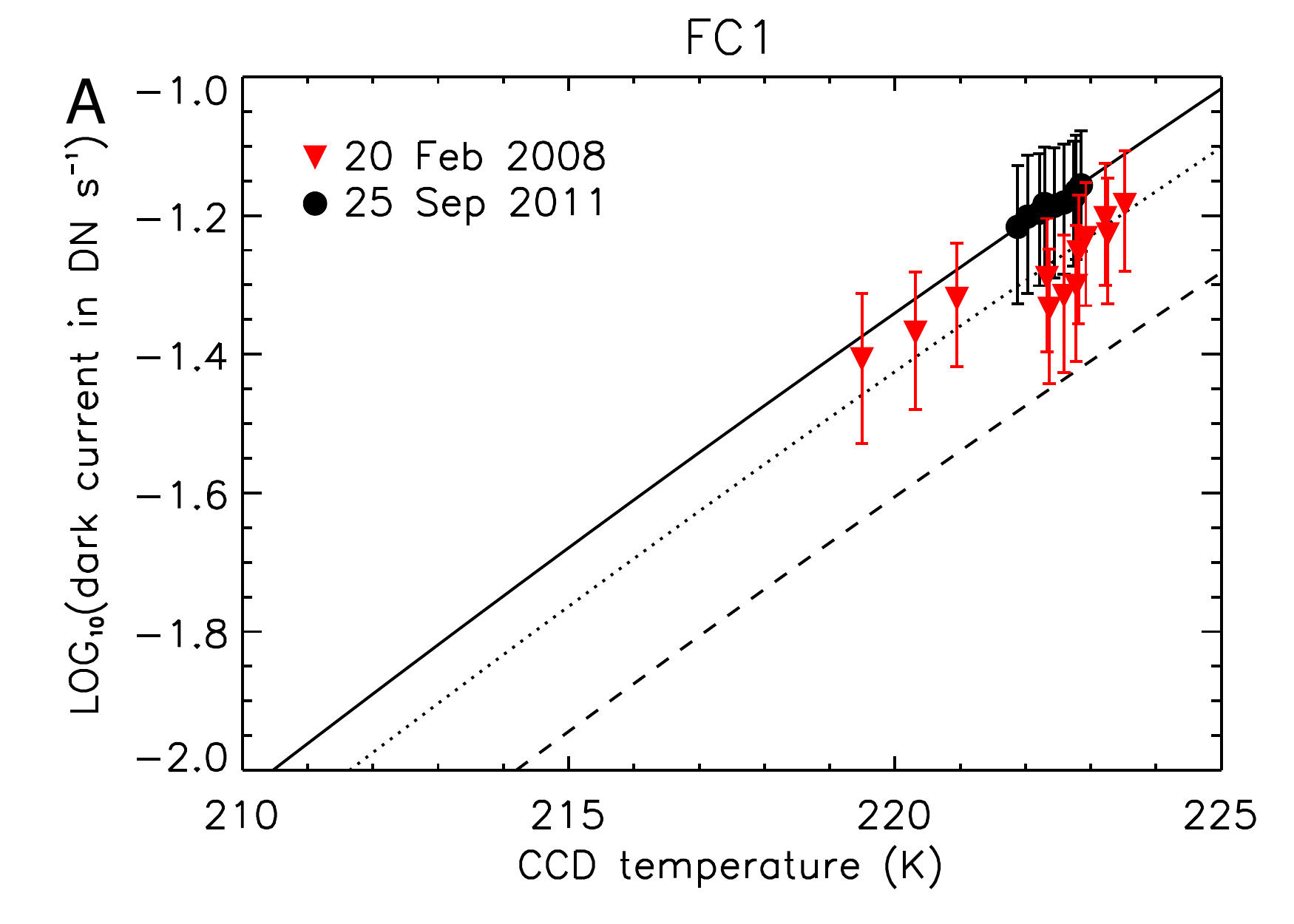}
\includegraphics[width=8cm,angle=0]{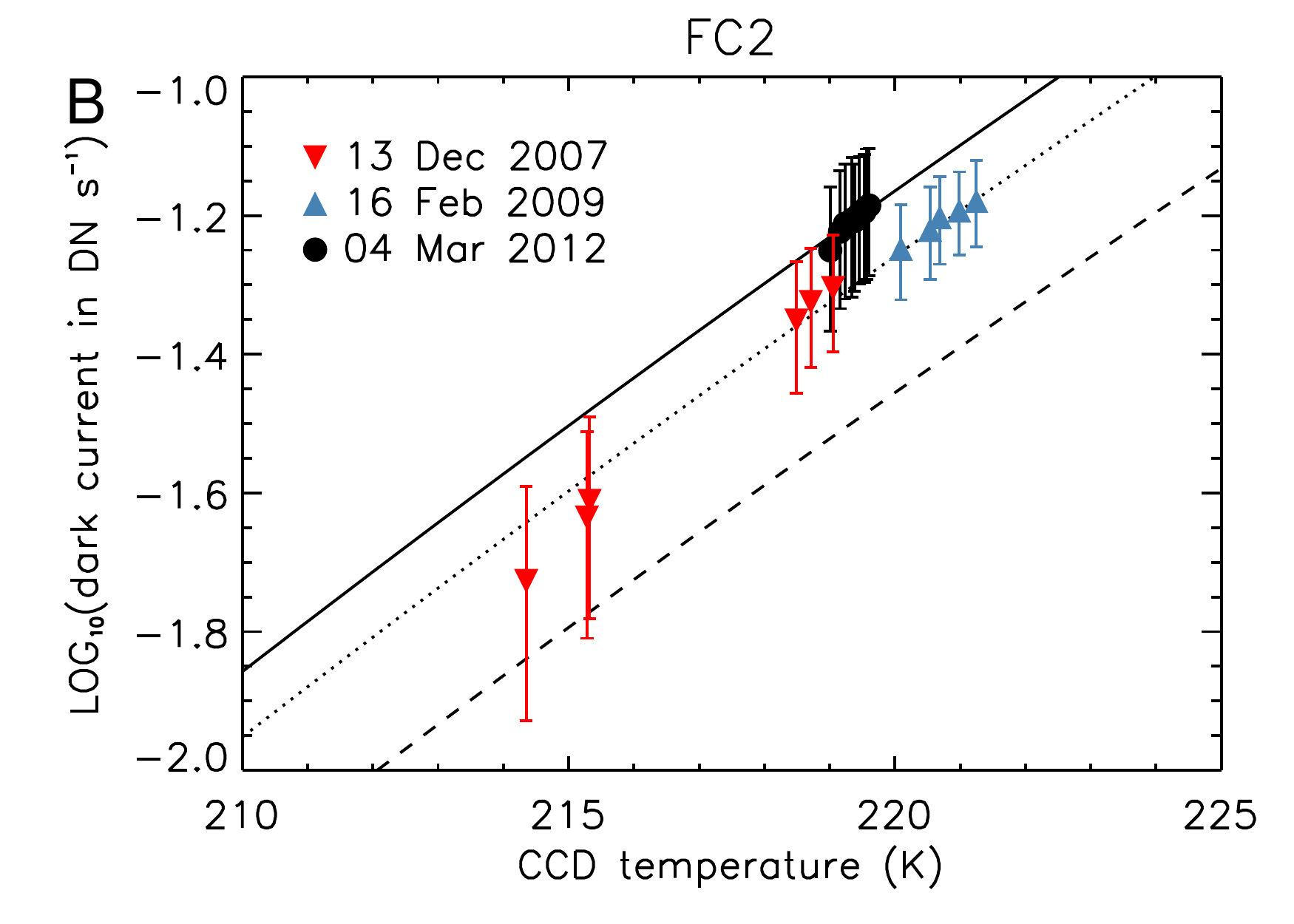}
\caption{Evolution of the dark current floor, estimated as the average of row 1000 of a dark exposure (excluding cosmic rays and hot pixels; error bars are the standard deviation). The lines are fits of the model in Eq.~\ref{eq:bulk_dark} with $b = 1.018\times10^{-19}$ m$^2$ kg s$^{-2}$. {\bf A}:~FC1. Shown are data for {\it ICO Performance \& Calibration} (2008) and {\it VTH} (2011). The fits have $a = 8.91\times10^{12}$ (pre-launch model; dashed line), $a = 1.35\times10^{13}$ (dotted line), and $a = 1.64\times10^{13}$ (solid line). {\bf B}:~FC2. Shown are data for {\it ICO Calibration} (2007), {\it MGA} (2009), and {\it LAMO} (orbit C13, 2012). The fits have $a = 1.26\times10^{13}$ (pre-launch model; dashed line), $a = 1.98\times10^{13}$ (dotted line), and $a = 2.46\times10^{13}$ (solid line). The {\it ICO} measurements have a higher (systematic) uncertainty than later ones as the image bias was returned in integer rather than the floating point format.}
\label{fig:bulk_dark}
\end{figure}

Residual charge was first identified in laboratory flat fields of FC1. The phenomenon has since been studied using a FC qualification model that is similarly affected. The residual charge pattern is highly variable from camera to camera, and difficult to display clearly, with one pixel showing hundreds of DN of residual charge, and the pixel next to it none. It is constant in appearance, but its strength depends on the incoming light flux in non-linear fashion. As such, it mostly affects clear filter images, with the exact pattern depending on the observed scene. The FC1 lab flat fields were acquired at relatively low flux. While residual charge is present in calibration lamp images, which are acquired in-flight typically twice per year, the lamp flux is also low. To characterize the residual charge at high light flux levels, FC1 images of Vesta were acquired during the {\it VTH} campaign, which took place during transfer from {\it Survey} to the {\it High Altitude Mapping Orbit} ({\it HAMO}). We analyze clear filter image {\bf 1241}, a 0.009~s exposure, to determine the residual charge contribution. Image {\bf 1242}, a zero-second exposure, was acquired immediately after {\bf 1241} and represents the $\mathbf{X} + \mathbf{S}$ terms in Eq.~\ref{eq:image}, because the charge rate $\mathbf{C}$ is the same in both images (ignoring any out-of-field stray light contribution). The bias is available for both images, and we can derive the clean image (including dark current) for this particular scene as
\begin{equation}
(\mathbf{C}_{1241} + \mathbf{D}_{1241}) t_{\rm exp} = (\mathbf{W}_{1241} - b_{1241}) - (\mathbf{W}_{1242} - b_{1242}).
\end{equation}
After constructing an artificially smeared image $\mathbf{S}_{1241}$ from this clean image (see Sec.~\ref{sec:smear}) and calculating the dark current contribution $\mathbf{D}_{1241}$, we derive the residual charge image $\mathbf{X}_{1241}$ through Eq.~\ref{eq:image}. The method described above only works because the scene hardly changed between images {\bf 1241} and {\bf 1242}. The charge rate of image {\bf 1241} is around 200~kDN~s$^{-1}$ (Fig.~\ref{fig:extra_charge}A). The histogram of $\mathbf{X}_{1241}$ (Fig.~\ref{fig:extra_charge}B) shows that a few pixels have a residual charge of close to 1000~DN, thousands of pixels have a residual charge around 300~DN, and hundreds of thousands of pixels around 100~DN. To put these numbers in perspective, we consider that the typical signal level at Vesta is 3-4~kDN, for which 300~DN of residual charge represents 7-9\%.

If ever it will prove necessary to switch to the backup camera FC1, image calibration should include the step of subtracting this charge. Dedicated observation campaigns will be required to better characterize the residual charge at the high flux levels expected at Ceres, and to establish a procedure to reliably extract $\mathbf{X}$ from $\mathbf{W}^i$.

\subsection{Read-out smear}
\label{sec:smear}

At the end of an exposure the image is transferred from the CCD active area to the covered storage area. The image is read-out from the storage area in little over one second \citep{S11}. During the rapid transfer (1.32~msec), the active area continues to be exposed. Consequently, the bottom rows will enter storage directly, but the top rows will accumulate charge all the way down to the storage area. This introduces a charge gradient from image top to bottom, which is known as the {\it electronic shutter effect}. If the exposure time is on the order of the shift time, the image needs to be corrected for this smear. This is generally the case for clear filter images of the asteroid surface.

\begin{figure}
\centering
\includegraphics[width=8cm,angle=0]{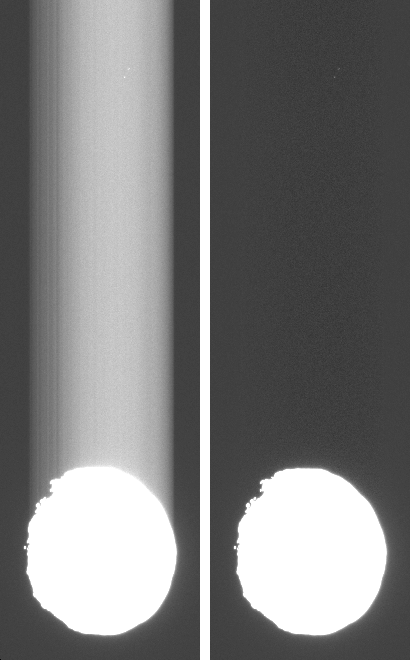}
\caption{Application of the read-out smear removal algorithm to {\it Rotational Characterization \#2} image {\bf 2770} of Vesta, shown here before ({\bf left}) and after ({\bf right}) removal. In both images the brightness is scaled such that black is $-50$~DN and white is 150~DN, which makes Vesta itself appear overexposed (it was not). The exposure time of this clear filter image was 8~msec, the signal on the Vesta disk itself is 3-5~kDN. While the average signal in the space background is $0.3 \pm 1.2$~DN, the lower brightness boundary was chosen negative to show that the algorithm removes the correct amount of signal.}
\label{fig:smear}
\end{figure}

At this stage of the calibration we have removed the bias and dark current from the raw image. For the moment, we assume that the stray light and residual charge contributions are zero, which leaves us with
\begin{equation}
\mathbf{W^\prime} = \mathbf{W} - \mathbf{D} t_{\rm exp} - b = \mathbf{C} t_{\rm exp} + \mathbf{S}(\mathbf{C}).
\end{equation}
We correct for the electronic shutter effect by means of an algorithm that assumes that the scene witnessed during the transfer is precisely that captured by the image. The smear depends on the clean image $\mathbf{C}$, which is unknown at this stage, so we estimate it from $\mathbf{W}^\prime$ in the following way. As it takes 1.32~ms to shift all 1056~rows of the full image frame, the shift time per row is $t_{\rm shift} = 1.25$~$\mu$s. Only 1024~rows (the active area) are actually exposed for a time $t_{\rm exp}$. The algorithm iteratively subtracts the image smear by calculating the smear contribution $S_y$ for row $y$ as
\begin{equation}
S_y = \frac{t_{\rm shift}}{t_{\rm exp}} W^\prime_y.
\end{equation}
The algorithm starts by calculating the smear $S_{16}$ for the bottom row of the active area, which is then subtracted from rows $\mathbf{W}^\prime_{17}$ to $\mathbf{W}^\prime_{1055}$, i.e.\ all rows above $\mathbf{W}^\prime_{16}$. This procedure is then repeated for row $\mathbf{W}^\prime_{17}$ (whose content has now changed!), and subsequently for the remaining rows of the active area ($\mathbf{W}^\prime_{18}$ to $\mathbf{W}^\prime_{1039}$). The result is the clean image $\mathbf{C} t_{\rm exp}$. A successful application of the read-out smear removal algorithm is shown in Fig.~\ref{fig:smear}. Note that this correction will fail for columns that contain saturated pixels, which is one of the reasons why the FC team was at pains to avoid overexposure at Vesta.

\subsection{Radiometric calibration}
\label{sec:rad_cal}

Devising a good radiometric calibration for the FC is a challenge because of the reduced pixel fill factor, due to the CCD being front-illuminated, and the strong in-field stray light in the narrow-band filters (see Sec.~\ref{sec:in-field} and \citealt{S11}). Prior to launch both cameras were thoroughly characterized in the laboratory. The transition to the space environment may have affected the radiometric response of the cameras. The in-flight observation of photometric standard stars and solar system objects allowed for a verification and refinement of the lab calibration.

\subsubsection{Methodology}
\label{sec:methodology}

The spectral responsivity, i.e.\ the detailed responsivity as a function of wavelength, of FC1 was determined on ground for each filter by having it observe the output of a monochromator through a diffuser and collimator (Fig.~39 in \citealt{S11}). Our {\it a priori} assumptions are that the FC1 detailed responsivities are also valid for FC2 (the cameras were built according to the same specifications), and that they have not changed in flight. The spectral responsivity was then used to calculate responsivity factors, which convert the DNs obtained through each filter into physical units. The factors of the ground calibration need to be revised in light of in-flight measurements.

First, we define some characteristics of the FC filters, as summarized in Table~\ref{tab:filters}. The filter effective wavelength (in nm) is calculated as
\begin{equation}
\lambda^i_{\rm eff} = \frac{\int_0^\infty \lambda r^i(\lambda) \mathrm{d}\lambda}{\int_0^\infty r^i(\lambda) \mathrm{d}\lambda},
\label{eq:eff_wav}
\end{equation}
with $r^i(\lambda)$ the spectral responsivity for filter $i$ in DN J$^{-1}$. Note that the filter central wavelength $\lambda_{\rm cen}$ in the table is defined as the average of the wavelengths that define the FWHM of the filter transmission curve. Likewise, the effective solar flux at 1~AU (in W m$^{-2}$ nm$^{-1}$) is defined as
\begin{equation}
F^i_\odot = \frac{\int_0^\infty F_\odot(\lambda) r^i(\lambda) \mathrm{d}\lambda}{\int_0^\infty r^i(\lambda) \mathrm{d}\lambda}.
\label{eq:sol_flux}
\end{equation}
For the solar spectrum $F_\odot(\lambda)$ we use the {\sc modtran} zero-air-mass solar flux\footnote{\url{http://rredc.nrel.gov/solar/spectra/am0/modtran.html}}.

The actual radiometric calibration is performed after the correction for image smear resulting from the electronic shutter effect, as described in the previous section. We divide the result of this step by the exposure time $t_{\rm exp}$ (in seconds), yielding the charge rate image $\mathbf{C}^i$ (in DN~s$^{-1}$). The image $\mathbf{L}^i$ in physical units of spectral radiance (W m$^{-2}$ nm$^{-1}$ sr$^{-1}$) is obtained by dividing the charge rate by the responsivity. For each pixel ($x,y$) we have:
\begin{equation}
L^i_{xy} = C^i_{xy} / R^i_{xy}.
\label{eq:rad_cal}
\end{equation}
$R^i_{xy}$ is an element of the responsivity image $\mathbf{R}$, which is the product of the filter responsivity factor $\mathcal{R}$ and the flat field $\mathbf{N}$:
\begin{equation}
\mathbf{R}^i = \mathcal{R}^i \mathbf{N}^i.
\end{equation}
The responsivity factors $\mathcal{R}^i$ are listed in Table~\ref{tab:filters} for each filter $i$. Flat field $\mathbf{N}^i$ is the ``true'' flat field for filter $i$, containing the pixel-to-pixel variations in responsivity. The narrow-band filter flat fields are not identical to the images of the inside of an integrating sphere obtained before launch, which are known to be affected by in-field stray light (see Sec.~\ref{sec:in-field}). For the calibration of standard star images we used flat fields that were ``flattened'' by dividing them by a best-fit polynomial surface. The expected standard deviation of the radiance as a consequence of photon noise is
\begin{equation}
(\sigma_L)_{xy}^i = \sqrt{\frac{L^i_{xy}}{t_{\rm exp} g R^i_{xy}}},
\label{eq:rad_stdev}
\end{equation}
with the detector gain $g = 17.7$ electrons per DN.

The conversion in Eq.~\ref{eq:rad_cal} is only valid in case the target is extended, as opposed to being a point source. The responsivity factor for filter $i$ (in J$^{-1}$ m$^2$ nm sr) is then calculated as
\begin{equation}
\mathcal{R}^i = \frac{A \Omega_{\rm px} c^i \Delta\lambda^i \int_{\lambda^i_{\rm lo}}^{\lambda^i_{\rm hi}} r^i(\lambda) F_\odot(\lambda) \mathrm{d}\lambda}{\int_{\lambda^i_{\rm lo}}^{\lambda^i_{\rm hi}} F_\odot(\lambda) \mathrm{d}\lambda},
\label{eq:responsivity}
\end{equation}
in which $A = 3.41\times10^{-4}$~m$^2$ is the FC aperture, $\Omega_{\rm px} = 8.69\times10^{-9}$~sr is the solid angle of a single pixel. The correction factors $c^i$ follow from a comparison of predictions from the ground calibration with results from in-flight observations (see below). Note that the responsivities in Eq.~\ref{eq:responsivity} are based on the assumption that the target of observation has a solar spectrum. They can be tuned, in iterative fashion, to match the spectrum of the target. Such an exercise for Vesta will have to be deferred to a follow-up paper. $\Delta\lambda^i$ is the FWHM of the transmission profile of filter $i$, listed in Table~\ref{tab:filters}. The integration boundaries $\lambda^i_{\rm lo}$ and $\lambda^i_{\rm hi}$ are the lower and upper boundaries of the FWHM. Unlike the narrow-band filters, the clear filter does not have a rectangular transmission profile. With the FWHM not clearly defined, $\Delta\lambda^1$ is not available. Being a broad- band, the intensity and reflectance associated with the clear filter are not uniquely defined, but depend on the spectrum of the target. For the color filters, calculating the radiance is meaningful if the bandwidth is narrow compared to the scale on which the reflectance spectrum exhibits significant changes. For the clear filter this is not the case. Therefore, for the definition of the clear filter responsivity factor we adopt Eq.~\ref{eq:responsivity} with the $\Delta\lambda^1$ term removed. The result, $\mathcal{R}^1$, has units of J$^{-1}$ m$^2$ sr. For the integration limits $\lambda^1_{\rm lo}$ and $\lambda^1_{\rm hi}$ we adopt 400 and 1100~nm, respectively, these approximately being the limits of the CCD responsivity.

If the target is an extended body like Vesta, radiance can be converted into the reflectance quantity {\it radiance factor} $r_{\rm F}$ \citep{H81}, which is the bidirectional reflectance of the surface relative to that of a Lambert surface illuminated normally, also known as ``I/F'':
\begin{equation}
(r_{\rm F})_{xy}^i = \pi d^2_{\rm V} L^i_{xy} / F^i_\odot,
\label{eq:reflectance}
\end{equation}
where $d_{\rm V}$ is the distance of Vesta to the Sun at the time of the observation (in AU). $F^i_\odot$ is the effective solar flux in Eq.~\ref{eq:sol_flux} for filters $i \in (2,8)$, listed in Table~\ref{tab:filters}.

\subsubsection{Standard stars}
\label{sec:standard_stars}

\begin{figure}
\centering
\includegraphics[width=12cm,angle=0]{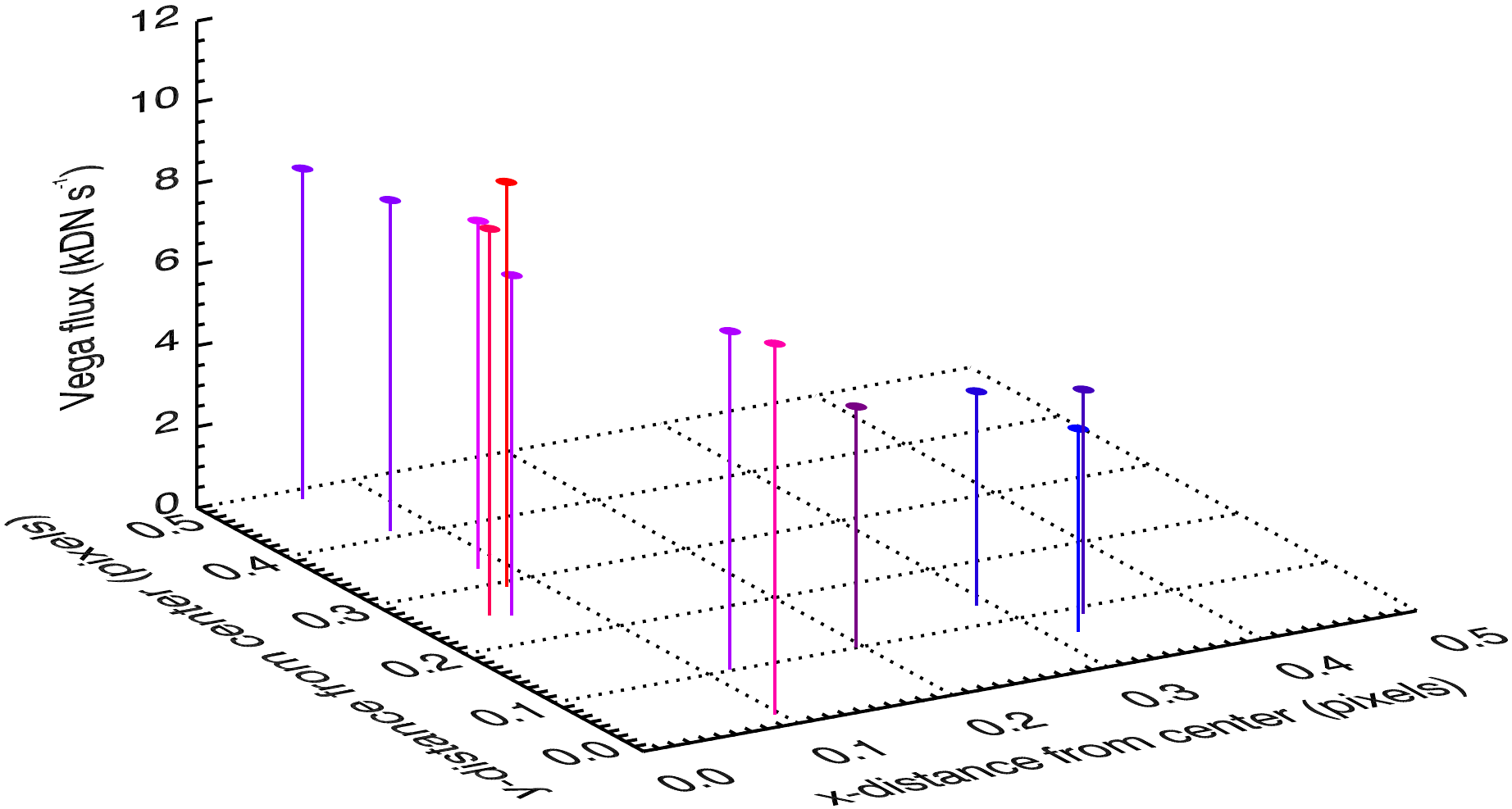}
\caption{The measured flux of a point source can depend strongly on the position of the PSF on the pixel for short exposure times, mostly in $x$-direction because of the presence of anti-blooming gates. The example shown here is the flux of Vega observed by FC1 through F7 during {\it ICO Performance} (0.91~s exposures) as a function of the distance of the center of the PSF to the center of a pixel, as determined by a 2D Gaussian fit. The data are colored on a scale ranging from blue (low) to red (high).}
\label{fig:fill_factor}
\end{figure}

In-flight observations were performed to verify the detailed spectral responsivities associated with FC1 that were determined in the laboratory before launch. During the {\it ICO Performance} and {\it Calibration} blocks both cameras observed several photometric standard stars (Sec.~\ref{sec:ICO}). Vega (spectral type A0V) was observed by both cameras during {\it Performance}. During {\it Calibration} FC1 and FC2 observed 73~Ceti (B9III) and 42~Pegasi (B8V), respectively. Stellar spectra in absolute flux units were retrieved from the European Southern Observatory website\footnote{\url{http://www.eso.org/sci/observing/tools/standards/spectra/}}. The availability of spectra in absolute units allows us to directly test the detailed responsivities $r^i(\lambda)$ by comparing the observed and expected flux in DN~s$^{-1}$. The outcome of this test are the correction factors in Eq.~\ref{eq:responsivity}. The expected flux $F_\star^i$ for filter~$i$ (in DN~s$^{-1}$) is calculated from the stellar spectrum $F_\star(\lambda)$ (in W m$^{-2}$ nm$^{-1}$) as
\begin{equation}
F_\star^i = A \int_0^\infty r^i(\lambda) F_\star(\lambda) \mathrm{d}\lambda.
\end{equation}
The observed flux (also in DN~s$^{-1}$) was derived from standard star images. These images were corrected for bias and dark current, and subsequently divided by a (flattened) flat field. They were not corrected for readout smear because the transfer time is negligible compared to the typical exposure time (1.32~ms versus tens of seconds). The flux was estimated as the total charge of a $15\times15$ pixel sized box centered on the star (determined from a 2D Gaussian fit), and corrected for background intensity (estimated as the median value of a $100\times100$ pixel sized area around the star). The box is large compared to the small FWHM of the PSF (1.0-1.6 pixels depending on filter, see \citealt{S11}), but chosen to include the broad wings. Vega is bright compared to 73~Cet and 42~Peg (apparent visual magnitude 0.03 versus 4.28 and 3.40, respectively), so exposures of Vega were relatively short, in the order of seconds. Anti-blooming gates and other obscuring structures on the surface of the CCD reduce the pixel fill factor \citep{S11}. Indeed, we find that the observed flux of the relatively short Vega exposures strongly depends on the position of the center of the PSF on the pixel (Fig.~\ref{fig:fill_factor}). For long exposures, however, tiny instabilities in the spacecraft pointing moves the stellar point spread function (PSF) over different parts of the pixel, effectively creating an average. To enable a meaningful comparison with the expected flux we must take the average of several observations through each filter. Such a sample is available for Vega ($n\sim10$), but 42~Peg and 73~Cet were observed only twice per filter.

\begin{figure}
\centering
\includegraphics[width=8cm,angle=0]{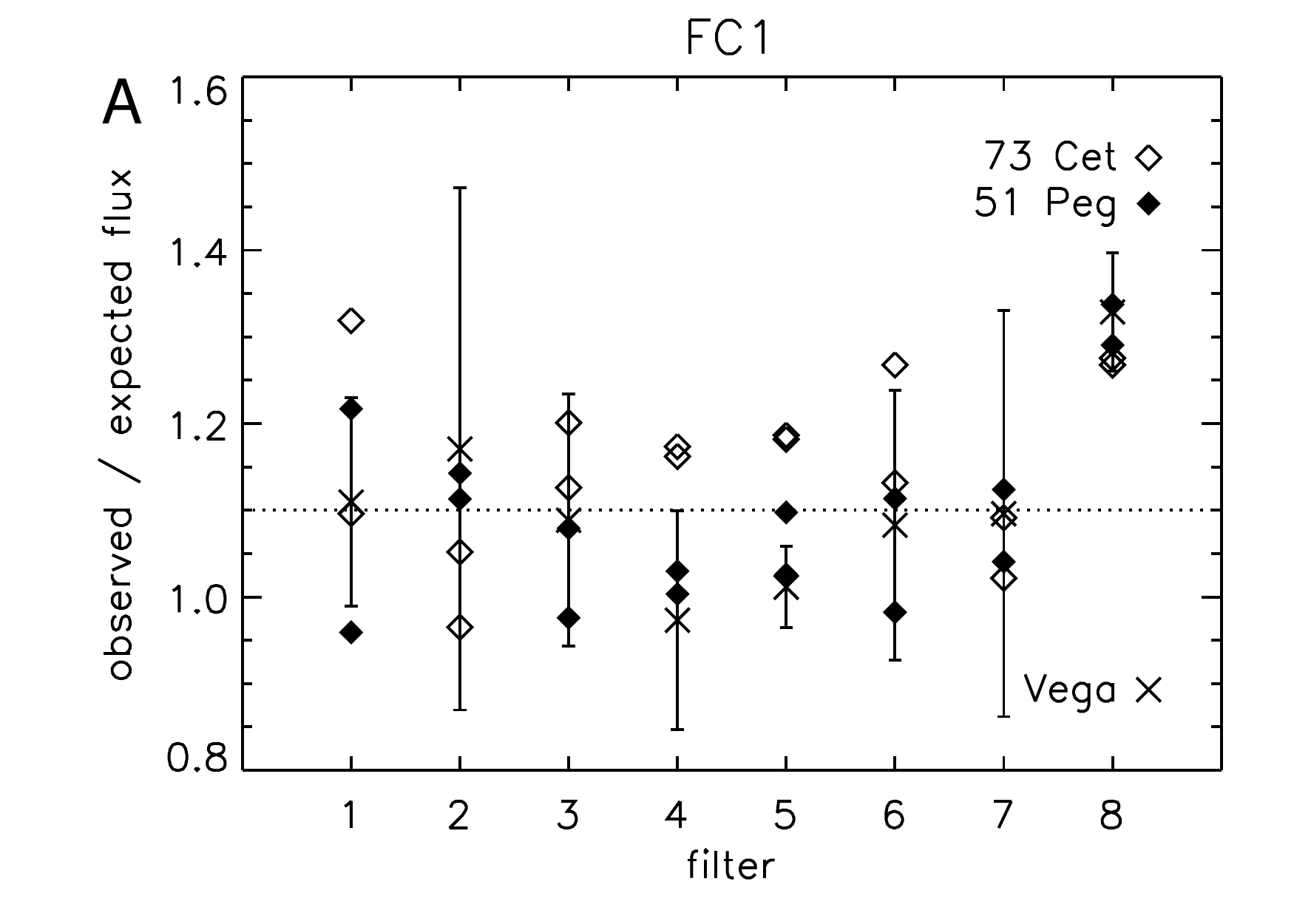}
\includegraphics[width=8cm,angle=0]{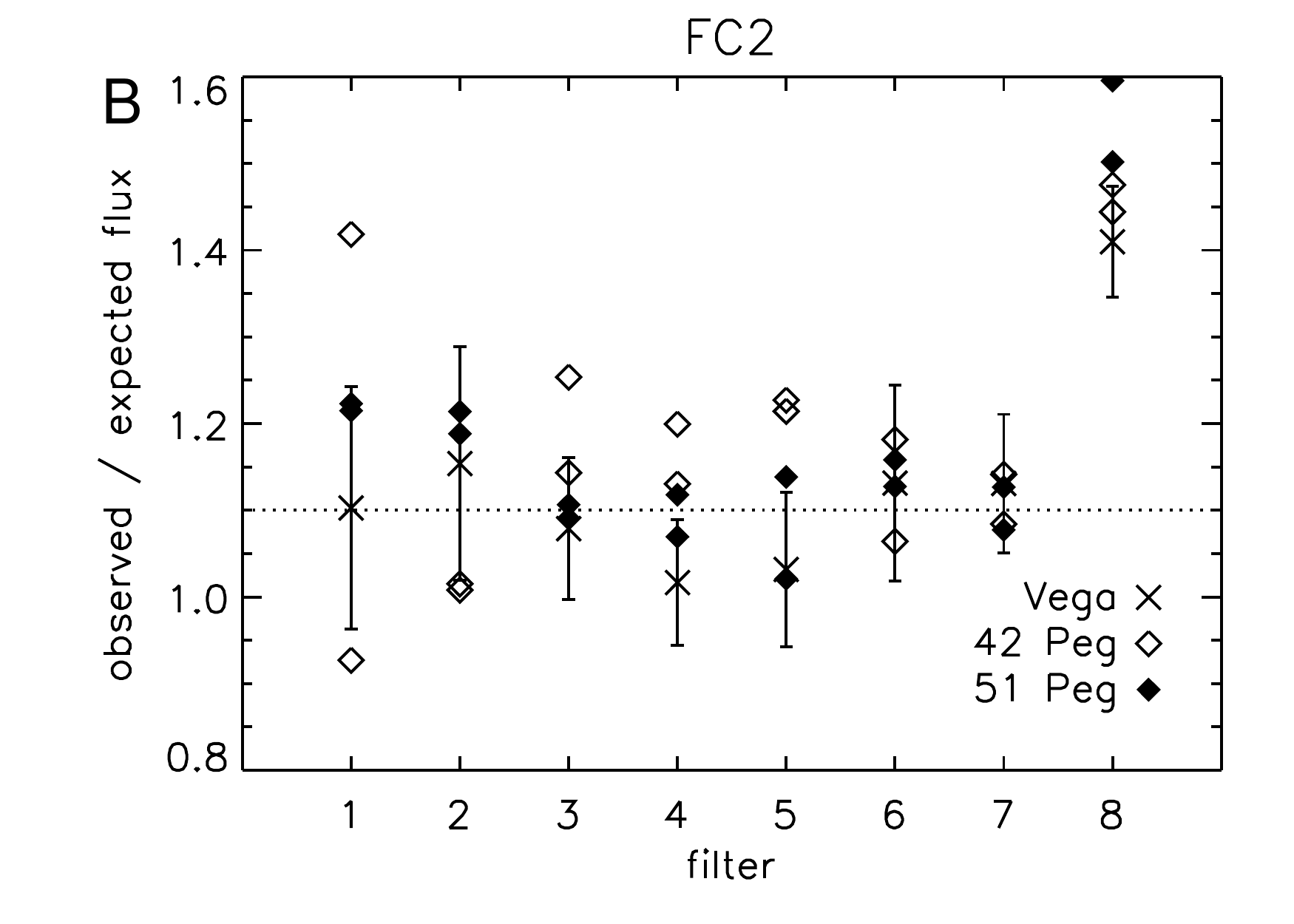}
\caption{The observed flux (in DN~s$^{-1}$) of photometric standard stars for all filters compared to that expected from the pre-launch radiometric calibration. The Vega data are the mean and standard deviation of $\sim$10 observations. 73~Cet, 42~Peg, and 51~Peg were observed twice per filter. The dotted line is shown for reference. The responsivities were calculated according to Eq.~\ref{eq:responsivity} with correction factors unity. {\bf A}.~FC1. {\bf B}.~FC2.}
\label{fig:star_flux}
\end{figure}

Figure~\ref{fig:star_flux} shows that the observed standard star flux (in DN~s$^{-1}$) is significantly higher for each filter than expected from the lab calibration. The large spread in the data is the result of the reduced pixel fill factor. The results for both cameras are very similar. The FC2 Vega data have a smaller standard deviation than the FC1 data, which suggests that FC1 pointing was more stable during the exposures. The average ratio of observed to expected flux in the clear filter (F1) is 1.11 for both cameras. It is around 1.1 for all color filters except F8, which has a ratio between 1.3 and 1.5. The F4 and F5 ratios are consistently higher (i.e. for both cameras) for 73~Cet and 42~Peg than for Vega. The spectrum of 73~Cet is similar to that of 42~Peg but both are significantly different from that of Vega in the F4 and F5 wavelength range; the local spectral slopes are similar, but the Vega flux is relatively high. A possible cause is that the spectral responsivity curves are slightly different in flight than determined on ground. However, while this may explain other irregularities in our results, in this case it is an unlikely explanation because the filter responsivity curves are relatively broad. This suggests that the ESO standard star spectra are not entirely accurate in the 900-1000~nm wavelength range, which may be due to the presence of atmospheric water absorption bands in that part of the spectrum. Other inaccuracies in the standard star spectra may lead to small inconsistencies in the results. The standard star observations reveal that the photometric response of both cameras is very similar. The full body of evidence suggests that the detailed laboratory responsivities are essentially correct, but off by about 10\%. The exception is filter F8, which is off by about 30\% for FC1 and 45\% for FC2. We infer the following correction factors: $c^1 = 1.11$, $c^{2\mhyphen7} = 1.10$, $c^8_{\rm FC1} = 1.30$, and $c^8_{\rm FC2} = 1.45$. The factors for F2-F7 are chosen slightly lower than that for F1 because the factor for F8 is so much higher than the others ($c^1$ includes all others).

\begin{figure}
\centering
\includegraphics[width=8cm,angle=0]{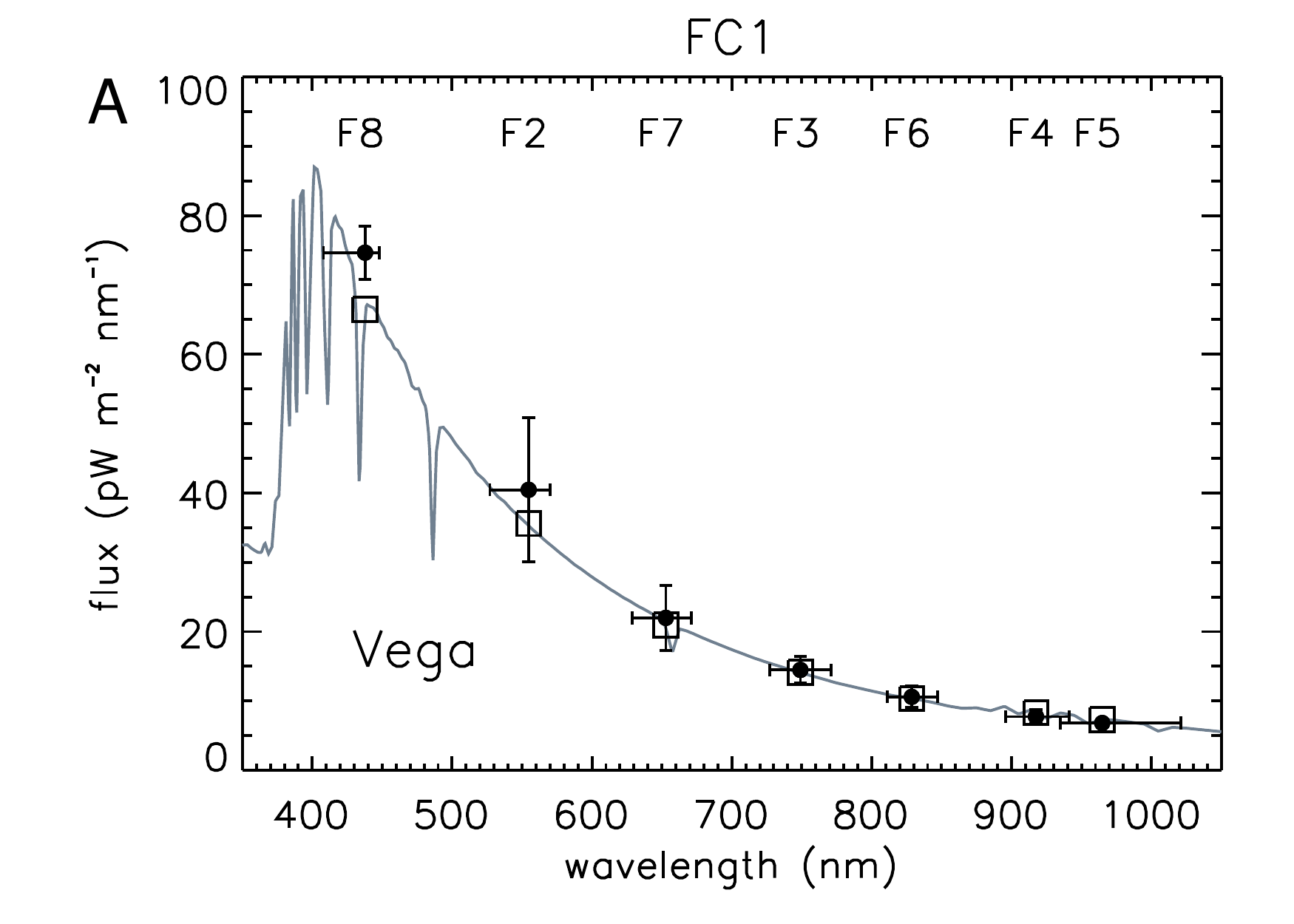}
\includegraphics[width=8cm,angle=0]{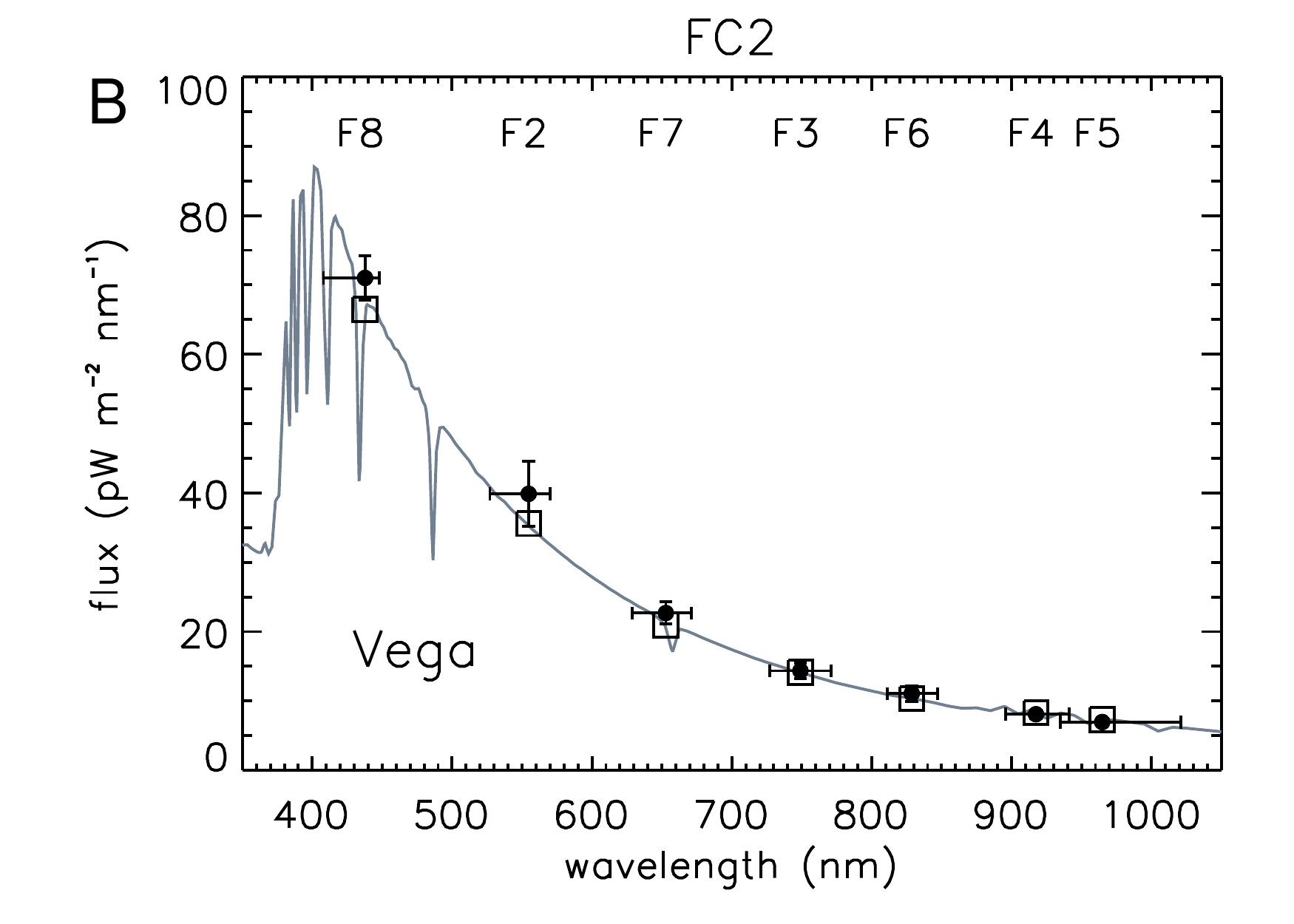}
\caption{Testing the Vega-specific responsivity factors by comparing the observed flux ($\bullet$) with the Vega spectrum using correction factors derived from Fig.~\ref{fig:star_flux} ($n = 10$). Also shown are the Vega flux averages in the filter passband ($\square$). Filter numbers are indicated. {\bf A}.~FC1. {\bf B}.~FC2.}
\label{fig:Vega}
\end{figure}

\begin{figure}
\centering
\includegraphics[width=8cm,angle=0]{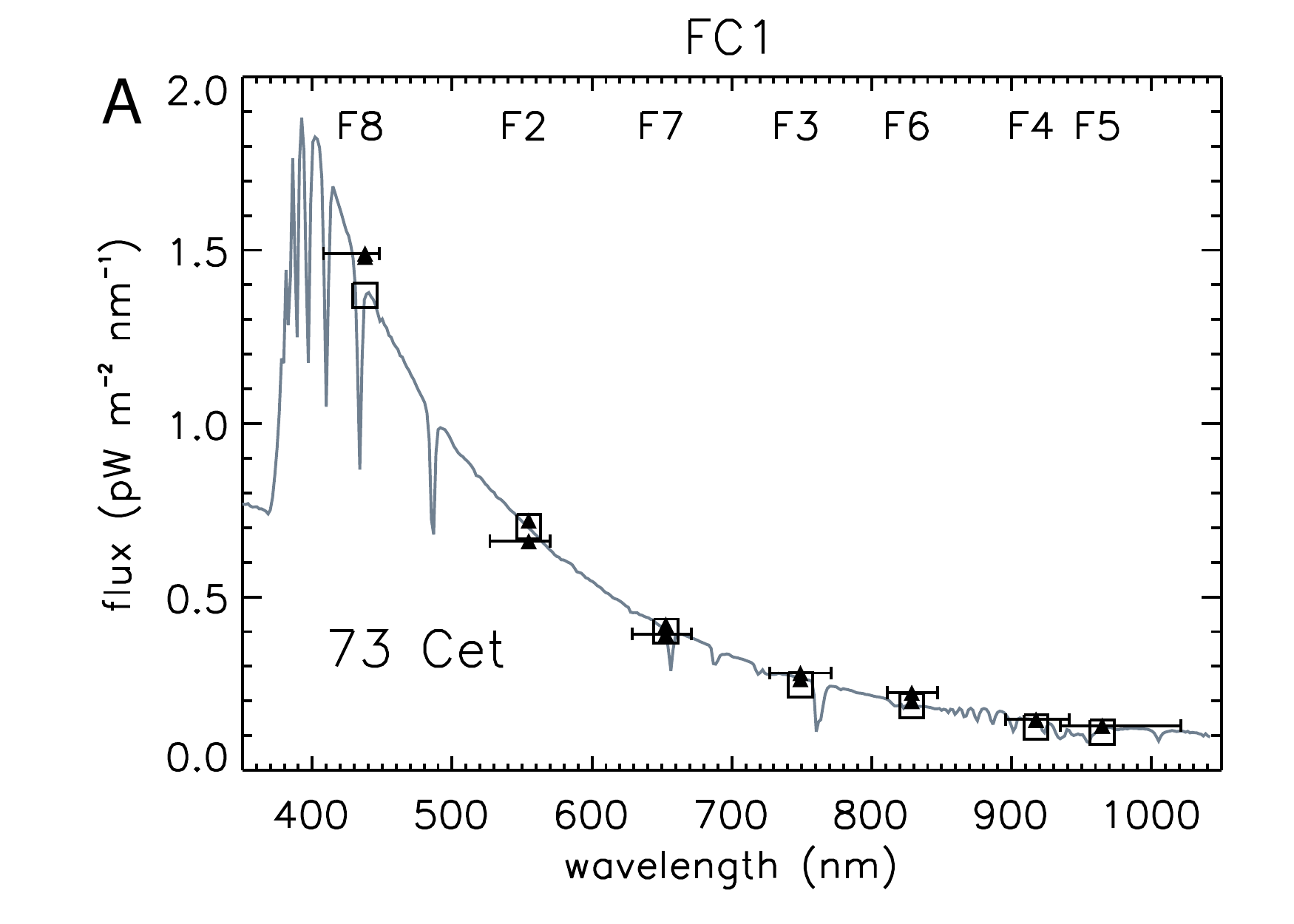}
\includegraphics[width=8cm,angle=0]{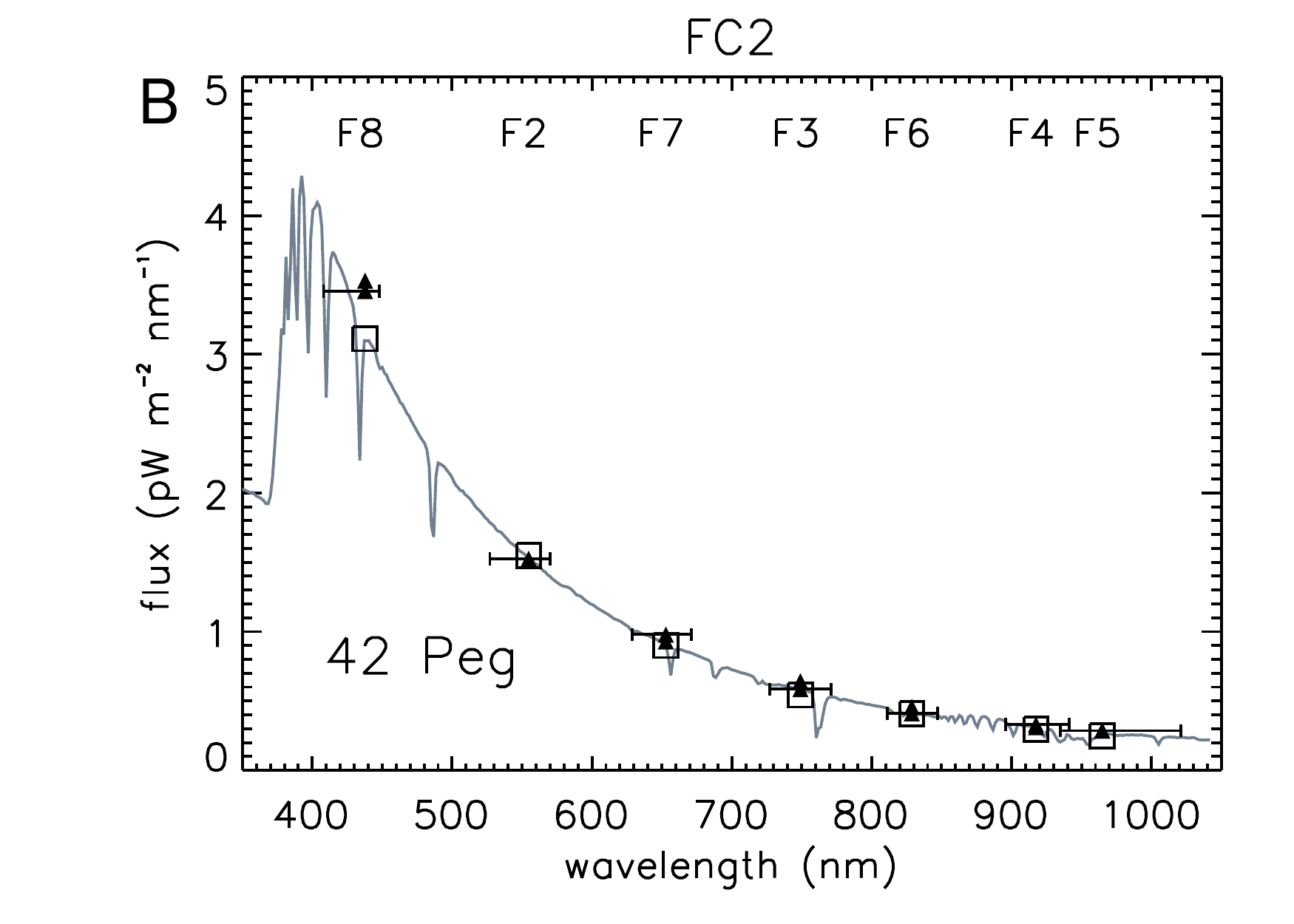}
\caption{Verification of the filter responsivities by comparing the observed integrated flux of standard stars ($\blacktriangle$) with the stellar spectrum ($n = 2$). The responsivities were calculated using correction factors and the stellar spectra instead of the solar spectrum. The FWHM is indicated only for one observation for clarity. Also shown are the stellar flux averages in the filter passband ($\square$). Filter numbers are indicated. {\bf A}.~FC1 observations of 73~Cet. {\bf B}.~FC2 observations of 42~Peg.}
\label{fig:Cet_Peg}
\end{figure}

Using these factors we can now test how well the color filter responsivity factors reconstruct the spectrum of Vega. We calculate responsivity factors according to Eq.~\ref{eq:responsivity}, but using the Vega spectrum instead of the solar spectrum. The observed stellar flux (in W m$^{-2}$ nm$^{-1}$) is obtained by calibrating the images using these revised responsivity factors $R^{i\prime}$ ($i = 2$-8), and integrating over a $15\times15$ sized box centered on the star:
\begin{equation}
F_\star^{i\prime} = \Omega_{\rm px} \sum C^i_{xy} / R^{i\prime}_{xy}.
\end{equation}
The results for both cameras are very similar, as shown in Fig.~\ref{fig:Vega}. The reconstructed flux for filters F8 and F2 appears to be slightly too high. We repeat the exercise for 73~Cet and 42~Peg (Fig.~\ref{fig:Cet_Peg}), and find that the reconstructions are accurate. The only exception is again filter F8, which is off by the same amount as for Vega. The reconstructions for F2 are accurate this time, which could be explained by inaccuracies in the ESO standard star spectra.

\subsubsection{Solar analog stars}

\begin{figure}
\centering
\includegraphics[width=8cm,angle=0]{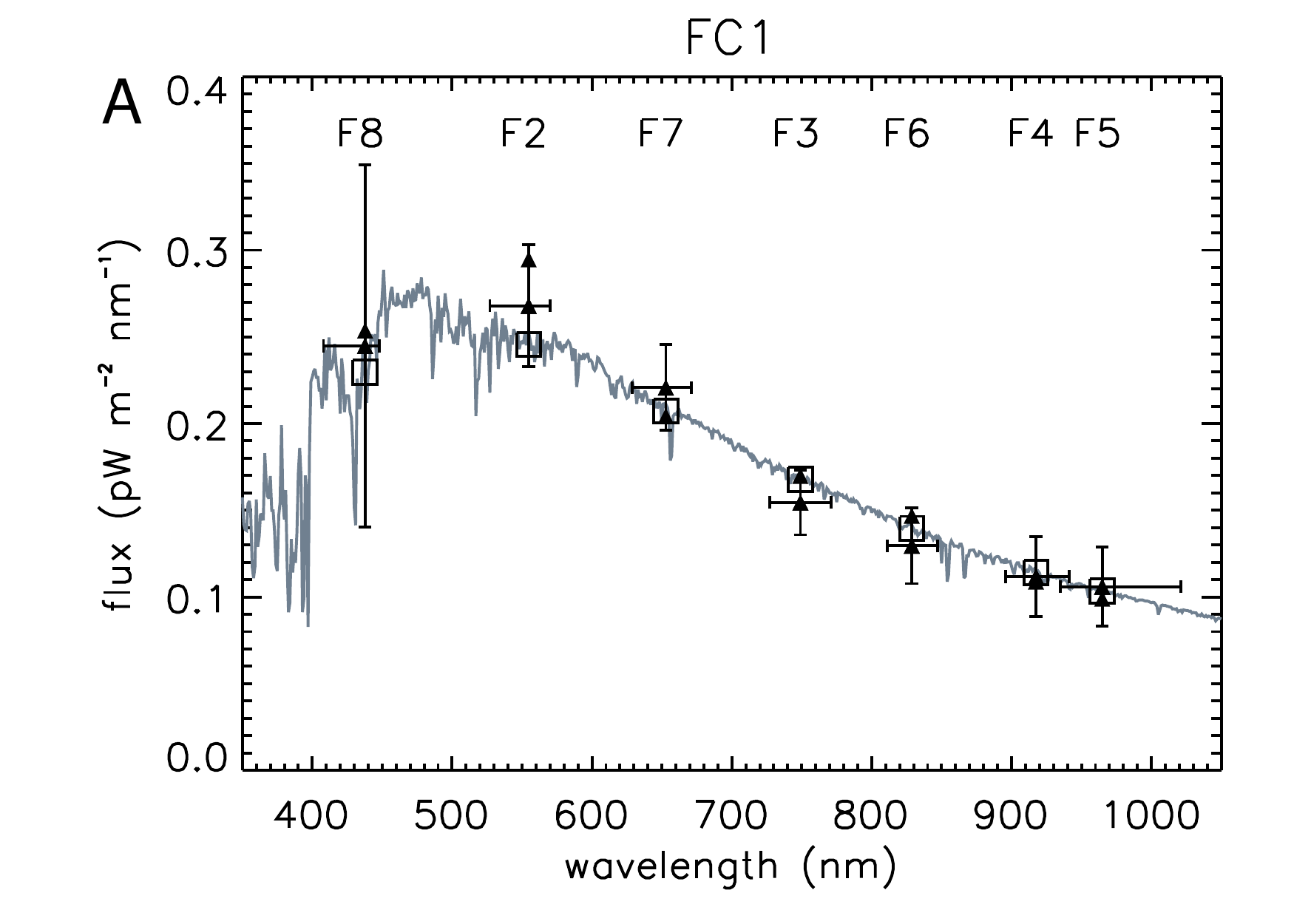}
\includegraphics[width=8cm,angle=0]{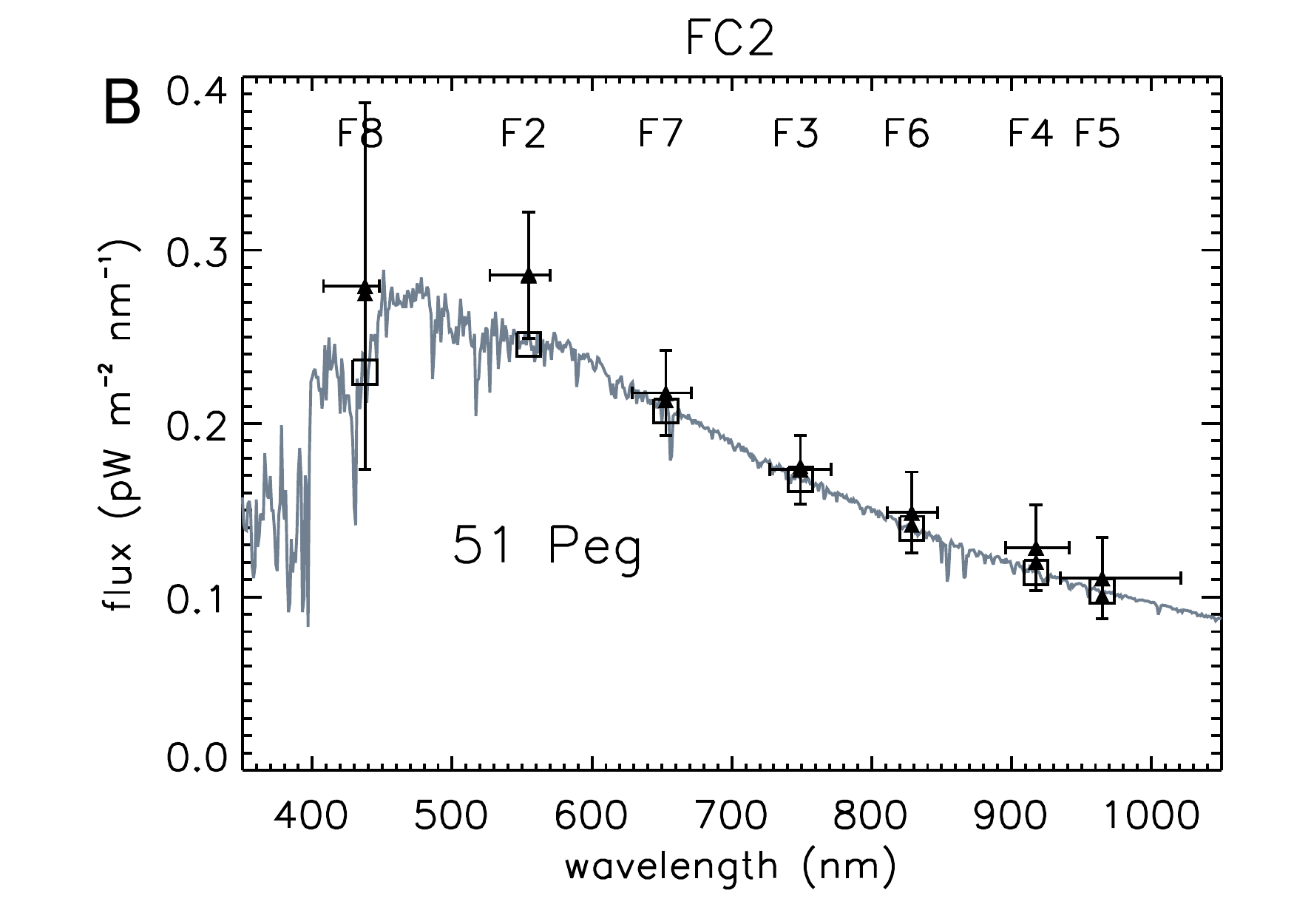}
\caption{Verification of the filter responsivities from FC2 {\it ICO Calibration} observations of solar analog 51~Peg. The responsivities were calculated using correction factors. The observed flux ($\blacktriangle$) was found by integrating the flux over a circle of 9 pixel radius surrounding the star. There were two observations in each filter; for clarity, error bars (photon noise) are shown for only one. For the spectrum of 51~Peg we adopt the solar spectrum, shown along with the averages in the filter passband ($\square$). Filter numbers are indicated. {\bf A}.~FC1. {\bf B}.~FC2.}
\label{fig:51 Peg}
\end{figure}

The photometric standard stars are biased towards early spectral types, ideal for accurately calibrating filters on the blue side of the spectrum, like F2 and F8. The flux of later type stars is more balanced over the camera wavelength range. This is why {\it ICO} included observations of a solar type star. Solar analog 51~Pegasi (G2.5IV) was observed twice per filter in the {\it Calibration} block. Being a solar analog, we adopt the solar spectrum for 51~Peg, converting it to absolute units by scaling it according to the apparent visual magnitudes of 51~Peg (5.46) and the Sun ($-26.74$). Because 51~Peg is relatively faint, the required exposure times were very long, which resulted in a more or less averaged signal over the pixel. Consequently, the flux reconstructed from individual images of 51~Peg does not show the large variability encountered for Vega. The comparison of the observed flux and the flux expected from the pre-launch calibration (in DN~s$^{-1}$) is included in Fig.~\ref{fig:star_flux}, and the results are broadly consistent with those for the photometric standard stars. In Fig.~\ref{fig:51 Peg} we evaluate the observed flux in physical units. Surprisingly, despite the long exposure times there is still considerable scatter in the data, especially for FC1. There are also significant differences between the observations of FC1 and FC2, indicating stable, but slightly different pointings for both models. Because of the low signal-to-noise, the flux reconstruction is hard to evaluate. All fluxes are consistent with the predicted spectrum within the error bars, but the fluxes in filters F8 (FC2) and F2 (FC1 and FC2) appear to be slightly overestimated.

\subsubsection{Saturn}

Saturn is not an ideal calibration target because it has rings. Not only do these have a different reflectance spectrum from the planet itself, but also the inclination angle at which we observe them can change considerably. Consequently, it is difficult to predict the brightness of Saturn as a function of phase angle. We compare FC2 observations with the Saturn albedo of \citet{K94}, which, at a phase angle of $2.7^\circ$, is essentially the geometric albedo. To make a quantitative comparison we introduce the ``effective radius'' of a disk equivalent to the illuminated surface of the planet-ring system, that varies with phase angle. The apparent diameter of the planet itself was 0.96 pixels at the time of observation. Aided by the 1-2 pixels wide PSF, the observations are effectively averages over the pixel, as opposed to the stellar point sources observed earlier.

\begin{figure}
\centering
\includegraphics[width=8.0cm,angle=0]{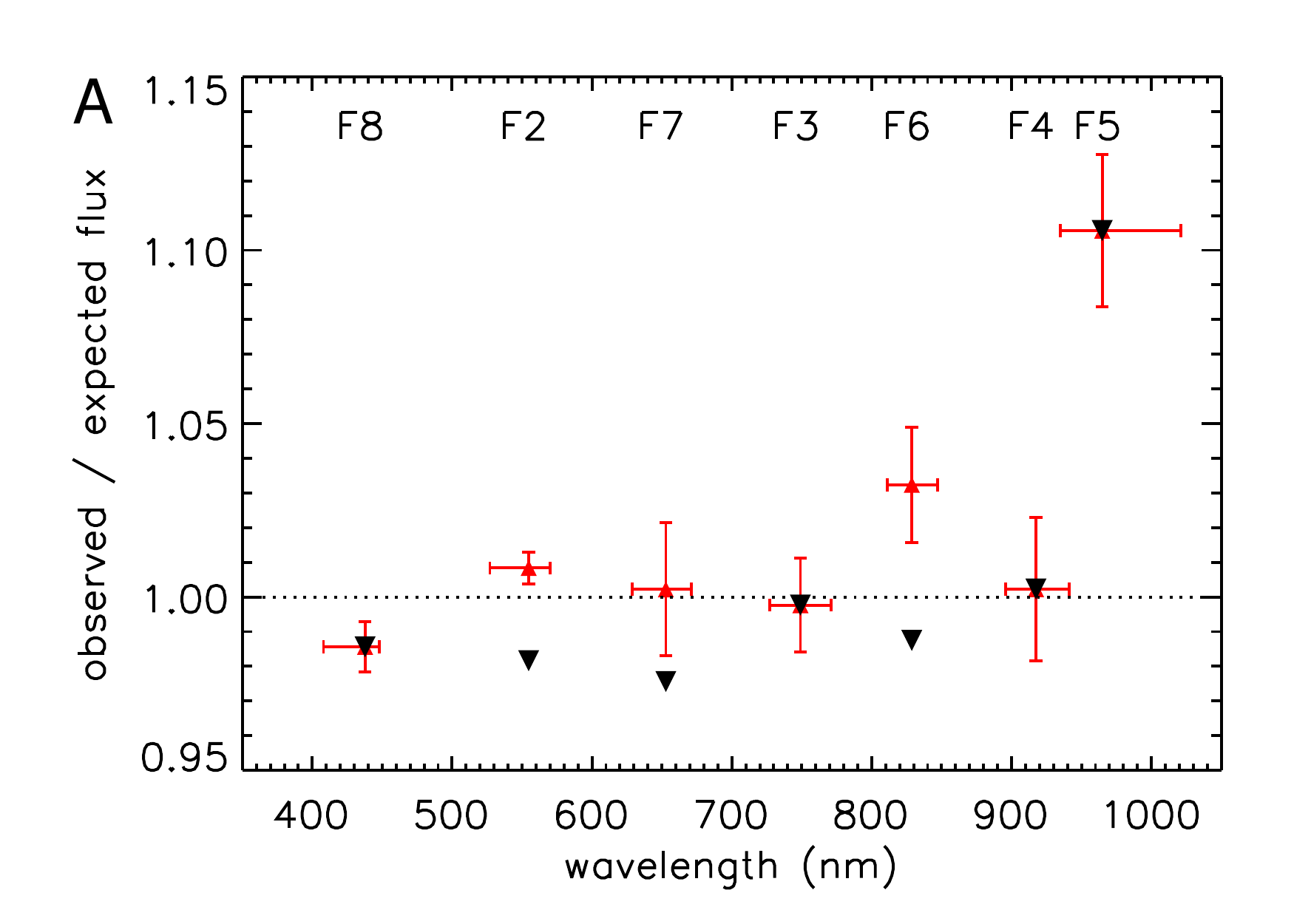}
\includegraphics[width=8.0cm,angle=0]{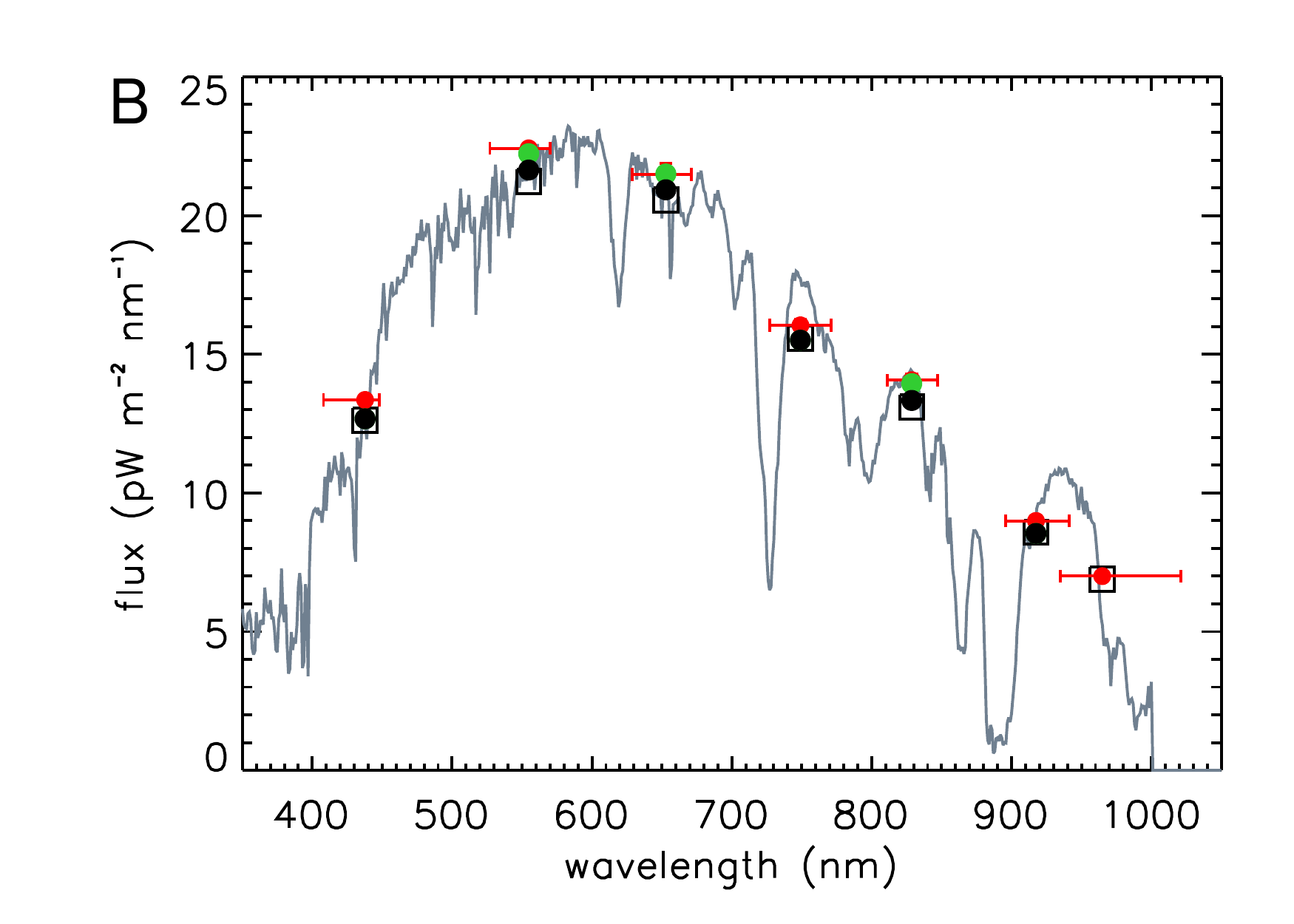}
\includegraphics[width=8.0cm,angle=0]{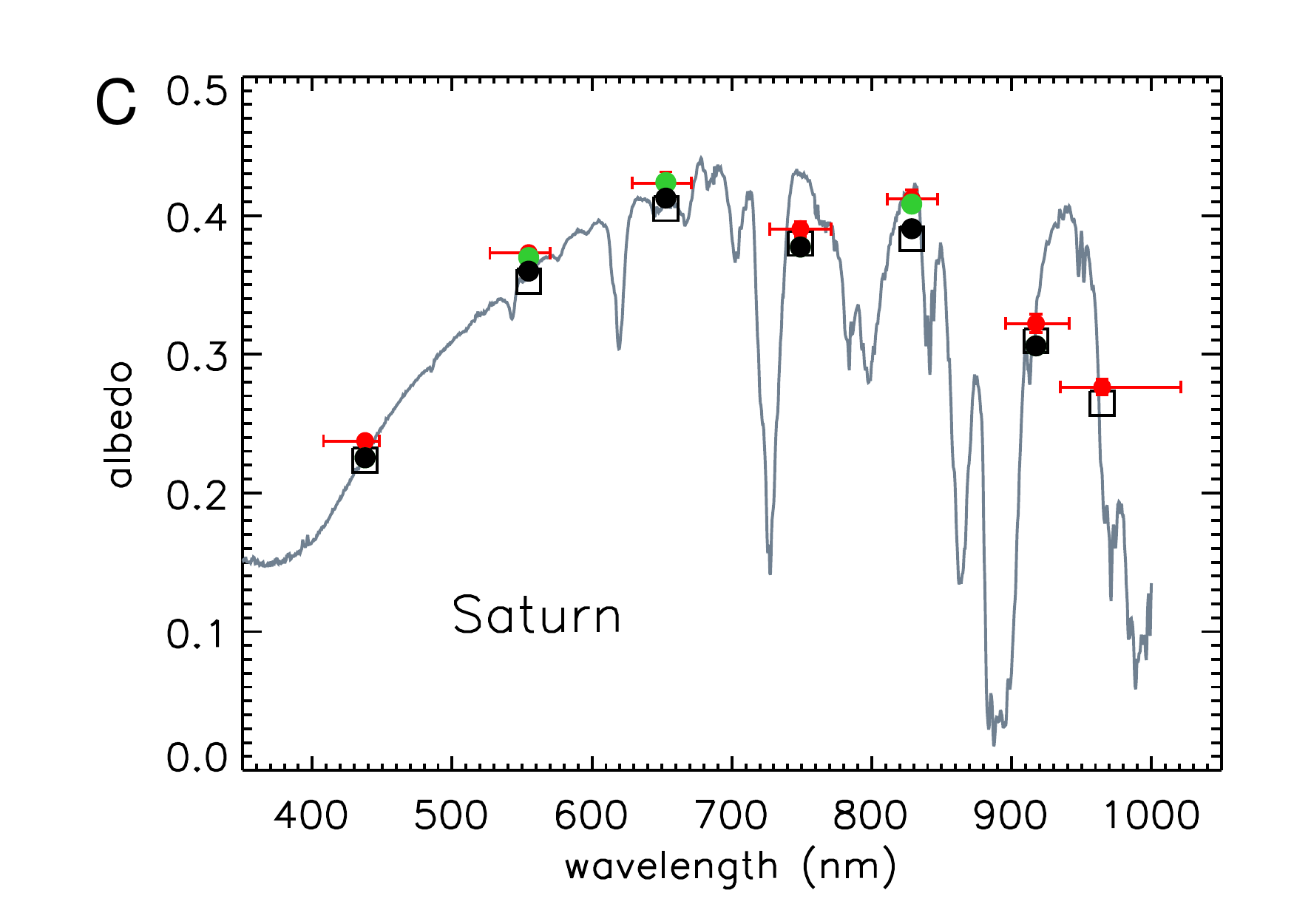}
\caption{Verification of the filter responsivities with FC2 {\it ICO Calibration} observations of Saturn. The intensity was integrated over a circle of 9 pixel radius surrounding the target ($n=8$). The Saturn albedo was adopted from \citet{K94}, assuming a disk radius of 60550~km, shown along with the averages in the filter passband ($\square$). For clarity, error bars are shown only for one of the symbol types. {\bf A}:~The observed flux (in DN~s$^{-1}$) compared to that expected, using the correction factors derived from standard star observations (\textcolor{red}{$\blacktriangle$}) and the revised factors in Table~\ref{tab:filters} ($\blacktriangledown$). The albedo beyond 1000~nm is not available (assumed zero), which underestimates the expected flux for filter F5. {\bf B}:~The observed flux (in pW m$^{-2}$ nm$^{-1}$) compared to that expected. (\textcolor{red}{$\bullet$}): Responsivities calculated for a target with a solar spectrum with the original correction factors. (\textcolor{LimeGreen}{$\bullet$}): Responsivities calculated for Saturn albedo with the original factors. ($\bullet$): Responsivities calculated for Saturn albedo with the revised factors. {\bf C}:~Reconstruction of Saturn's albedo through Eq.~\ref{eq:reflectance}. Plot symbols as for (B).}
\label{fig:Saturn}
\end{figure}

First we test the detailed responsivities corrected by the factors derived from the standard star observations. Figure~\ref{fig:Saturn}A compares the observed flux with that expected (in DN~s$^{-1}$). The single free parameter, the effective radius of Saturn, is chosen such that we obtain ratios closest to unity for filters F3 and F7, for which we consider the responsivity factors most reliable (based on the standard star observations). The flux in filter F4 is also very close to that expected, but the other filters appear to be off by various degrees. The flux in filter F6 is too high by about 3\%, whereas the F2 and F8 fluxes are too high and too low, respectively, by around 1\%. Note that this is not a good test for filter F5; the albedo beyond 1000~nm is not part of the \citet{K94} data set and assumed to be zero, which leads to underestimating the expected flux. The estimated effective radius of 60550~km for our observations at phase angle $7.0^\circ$ compares well to the equatorial radius of the planet of 60270~km, and includes the ring contribution.

Next, we compare the observed and expected fluxes in physical units (W m$^{-2}$ nm$^{-1}$) in Fig.~\ref{fig:Saturn}B. The expected flux in this case is the Saturn albedo multiplied by the solar spectrum. Adopting the effective radius from Fig.~\ref{fig:Saturn}A, this test has no free parameters. We provide two reconstructions, one with the responsivity factors calculated for a target with the solar spectrum (red symbols), and one customized for the Saturn spectrum (green symbols), both adopting the correction factors derived from the standard star observations. Both versions appear to reconstruct the Saturn spectrum well, the customized responsivities slightly more accurately. It does appear, however, that the F2, F7, and F6 fluxes are slightly overestimated.

A final test is that of our ability to reconstruct the reflectance through Eq.~\ref{eq:reflectance}. The comparison of the reconstruction with the \citet{K94} albedo in Fig.~\ref{fig:Saturn}C is more straightforward than in Fig.~\ref{fig:Saturn}B, especially at shorter wavelengths, due to the absence of solar absorption lines. It shows more clearly that the albedos for F2, F7, and F6 are indeed overestimated by a few percent (red and green symbols). However, the albedo reconstruction for F8 is accurate for the customized responsivity factor. As a further, minor, improvement we can calculate the filter effective wavelength using the detailed responsivity multiplied with the observed spectrum as weight factor, rather than simply the detailed responsivity (Eq.~\ref{eq:eff_wav}). This makes no perceptible difference for any of the filters except F8, for which the effective wavelength shifts from 438 to 439~nm.

As calibration target, we attach more weight to Saturn than the standard stars because it is essentially an extended object, not affected by the reduced pixel fill factor. We therefore revise the correction factors for filters F2, F6, and F7 to the values in Table~\ref{tab:filters}. These are corrections to the factors converting the raw data numbers into physical units. Despite the inaccuracies observed for standard star observations through filter F8, the Saturn observations prove to be accurate. Because the standard star and Saturn spectra have opposite slopes in the F8 passband, and the detailed responsivity curve of F8 is strongly skewed, this inconsistency hint at inaccuracies in the latter. The new responsivity factors (with the F8 factors unchanged) are listed in Table~\ref{tab:filters}. The consequences of this revision for the reconstruction of Saturn's spectrum are also shown in Fig.~\ref{fig:Saturn} (black symbols). All reconstructed albedos calculated with the new factors are consistent with the 2\% accuracy quoted by \citet{K94}.

\section{Geometric distortion}
\label{sec:geom_dist}

The degree of geometric image distortion has been determined from images from the {\it ICO Performance} and {\it Calibration} blocks that were pointed at star fields. We adopt the distortion model described by \citet{HS97}. Let $(x_{\rm u}, y_{\rm u})$ be the (undistorted) CCD coordinates of a star (measured in mm from the center of the CCD) that would result from an idealized pinhole camera projection. Then
\begin{equation}
\begin{bmatrix}
x_{\rm u} \\
y_{\rm u}
\end{bmatrix} = f^i
\begin{bmatrix}
\tan \delta_x \\
\tan \delta_y
\end{bmatrix},
\label{eq:undistorted}
\end{equation}
with $f^i$ the focal length for filter $i$ (in mm), and $\delta$ the angle at which the star is observed in the sky (in radians). The true (distorted) horizontal CCD coordinates $(x_{\rm d}, y_{\rm d})$ in mm are different due to radial and tangential distortion:
\begin{equation}
\begin{bmatrix}
x_{\rm d} \\
y_{\rm d}
\end{bmatrix} =
\begin{bmatrix}
x_{\rm u} \\
y_{\rm u}
\end{bmatrix}
(1 + k_1^i r^2)+
\begin{bmatrix}
2 p^i_1 x_{\rm u} y_{\rm u} + p^i_2 (r^2 + 2 x_{\rm u}^2) \\
2 p^i_2 x_{\rm u} y_{\rm u} + p^i_1 (r^2 + 2 y_{\rm u}^2)
\end{bmatrix},
\end{equation}
with $r^2 = x_{\rm u}^2 + y_{\rm u}^2$. The distance coordinates $(x_{\rm d}, y_{\rm d})$ are translated into pixel coordinates $(u, v)$ as follows:
\begin{equation}
\begin{bmatrix}
u \\
v
\end{bmatrix} =
\begin{bmatrix}
c_x & 0 \\
0 & c_y
\end{bmatrix}
\begin{bmatrix}
x_{\rm d} \\
y_{\rm d}
\end{bmatrix}+
\begin{bmatrix}
u_0 \\
v_0
\end{bmatrix},
\label{eq:pixel_coord}
\end{equation}
with coefficients $(c_x, c_y)$ in units of mm$^{-1}$, and $(u_0, v_0) = (511.5, 511.5)$ the coordinates of the center of the CCD in pixels. The coefficients $(c_x, c_y)$ map millimeters to pixels in the focal plane $x$- and $y$-directions, and are identical to the inverse of the pixel size. If $c_x = c_y$ then the pixels are square, but for now we leave the option open that they are not. Thus, we use a 5-parameter model for image distortion in each filter $i$: the focal length $f^i$, the radial distortion parameter $k^i_1$, the tangential distortion parameters $p^i_1$ and $p^i_2$, and $c_y / c_x$. We verified that the optical axis coincides with the center of the image.

\begin{figure}
\centering
\includegraphics[width=8cm,angle=0]{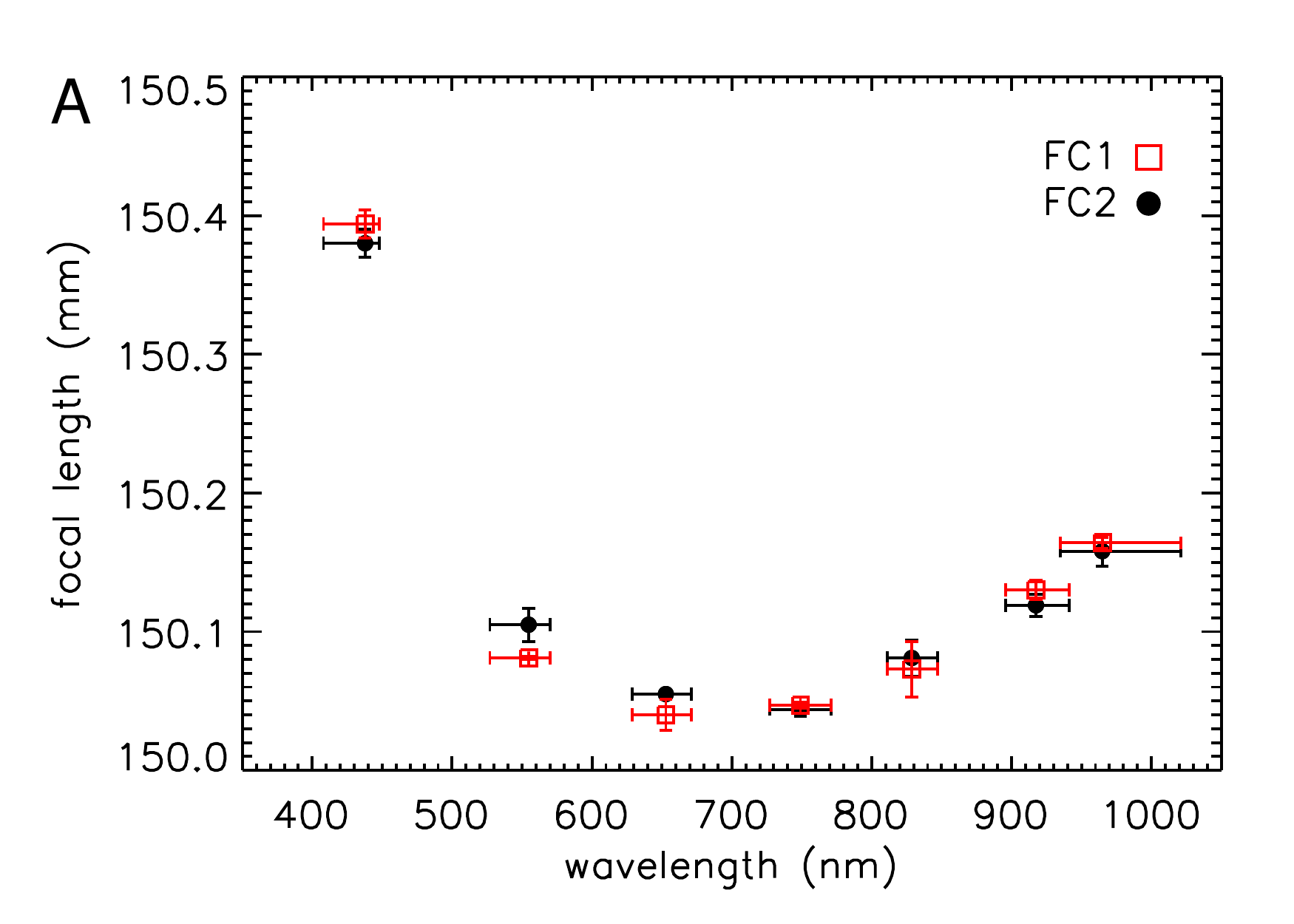}
\includegraphics[width=8cm,angle=0]{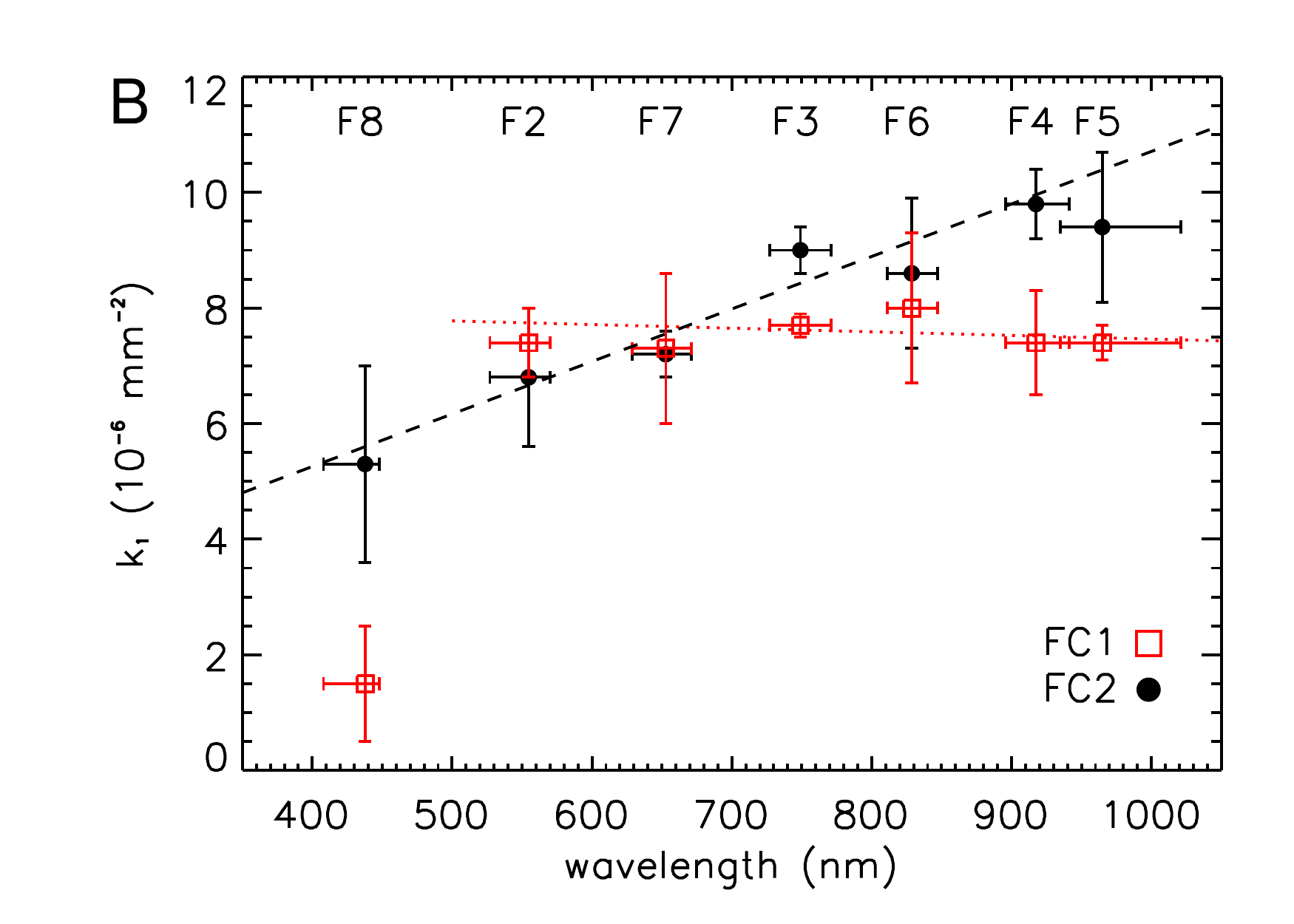}
\caption{FC geometric distortion characteristics (data in Table~\ref{tab:geom_dist}). {\bf A}:~The focal length varies with wavelength in the same way for FC1 and FC2. {\bf B}:~The FC2 radial distortion parameter $k_1$ depends on wavelength (lateral chromatic aberration), whereas that of FC1 is approximately constant, except for F8. The lines are linear fits to the data. The fit for FC2 is $k_1 = 1.62 + 0.00909 \lambda$, with $k_1$ in units of $10^{-6}$~mm$^{-2}$ and $\lambda$ in nm. The fit for FC1 is $k_1 = 8.09 - 0.000628 \lambda$.}
\label{fig:geom_dist}
\end{figure}

The star field around 20~Cephei was imaged during the {\it ICO Performance} and {\it DC041} campaigns (Sec.~\ref{sec:ICO} and \ref{sec:semi-annuals}). For 149 stars in this field we retrieved the International Celestial Reference System (ICRS) sky coordinates (in R.A. and Dec.), and fitted our model to the observed positions, estimated by means of a 2D Gaussian fit to the stellar brightness profiles. We find the FC focal length to be around 150.1~mm, the exact value depending on the filter (Table~\ref{tab:geom_dist} and Fig.~\ref{fig:geom_dist}A). The focal length values are in very good agreement with pre-launch measurements. With respect to the focal length, there are no significant differences between FC1 and FC2. However, we find that the optical properties are significantly different in the horizontal ($x$) and the vertical ($y$) direction, albeit by a very small amount (which is the same for both cameras). This implies that either the focal length is different in the $x$ and the $y$ direction, or that the pixels are non-square. From the data we are unable to distinguish between these two options, as from the model perspective they are equivalent. Thus, from a practical viewpoint it is irrelevant which of the two options corresponds to reality. We know that the physical pixels are square to one part in a thousand (the manufacturer lists the CCD dimensions as $14.34\times14.34$~mm$^2$ for $1024\times1024$ pixels), but the suspected non-squareness is on a scale smaller than that. Here, we assume that the width of the CCD in the $x$-direction is 14.340~mm, which corresponds to a pixel width of 14.004~$\mu$m ($c_x = 71.409$~mm$^{-1}$). From the star field observations we estimate $c_y / c_x = 1.00063\pm0.00003$ (averaged over all filters of both cameras), which results in the IFOV values listed in Table~\ref{tab:geom_dist}. The camera FOV is approximately $5.47^\circ$ squared.

Including radial distortion in the pinhole camera model typically reduced the residuals of the fit to the star positions by 10-15\%. Only the first order radial distortion parameter ($k_1$) is significantly different from zero. For all filters $k_1$ is found to be larger than zero, which implies that the FC suffers from slight pincushion distortion, amounting to half a pixel in the image corners. Including tangential distortion improved the fit by a further 5\%, but the $p_1$ and $p_2$ parameters are badly constrained, and were found to vary from image to image. The final fit results in Table~\ref{tab:geom_dist} were obtained by assuming zero tangential distortion. With respect to radial distortion, the two cameras are subtly different (Fig.~\ref{fig:geom_dist}B). For FC1, $k_1$ is essentially the same for all filters except F8. For FC2, $k_1$ linearly depends on wavelength, a phenomenon known as lateral chromatic aberration (Fig.~\ref{fig:geom_dist}). For this camera we adopt the $k_1$ values that result from a linear fit to the data (Table~\ref{tab:geom_dist}). The linear fit for all filters of FC1, except F8, has a slope very close to zero, which leads us to assume that $k_1$ is constant (the tabulated vales are the filter average). The final residuals of the fit of the model in Eqs.~\ref{eq:undistorted}-\ref{eq:pixel_coord} and the parameters in Table~\ref{tab:geom_dist} are typically around 0.1 pixel, and smaller than 0.3 pixel for almost all stars. For example, the RMS in the $x$- and $y$-direction are 0.091 and 0.110 pixels for a 15~s F1 exposure during {\it DC041} ($n$ = 87). The residuals may partly result from the inability of the Gaussian fit algorithm to find the true center of the stellar brightness profile due to the reduced pixel fill factor. We verified that the FC2 focal lengths and distortion parameters did not change in the period between {\it ICO Performance} and {\it DC041}.

In conclusion, the focal lengths have been retrieved with high accuracy and differences between the filters can play a dominant role in the geometric distortion. For example, images corrected for geometric distortion acquired in filters F3 and F8 differ in size by about one pixel in the image corners. Not correcting color composites of the sky for differences in focal length leads to noticeable color separation for stars. Radial (pincushion) distortion is small but significant, and $k_1$ parameters were retrieved with reasonable accuracy for each filter. Tangential distortion is insignificant. Consequently, the $p_1$ and $p_2$ parameters could not be estimated reliably, and are assumed to be zero. The FC geometric distortion parameters are documented in the {\sc spice} instrument kernel\footnote{The latest FC instrument kernel at the time of writing is \texttt{dawn\_fc\_v03.ti}.}.

\section{Stray light}
\label{sec:stray_light}

There are two sources of stray light that can affect image quality. Out-of-field stray light may result from light sources outside the FOV. The FC design was aimed at minimizing this type of stray light, but it still needs to be considered when searching for moons (the light source is the asteroid itself) or dust (observations at high solar phase angle). In-field stray light results from light sources inside the FOV, i.e.\ the scene imaged. It is significant for the FC narrow-band filters.

\subsection{In-field}
\label{sec:in-field}

\begin{figure}
\centering
\includegraphics[width=9cm,angle=0]{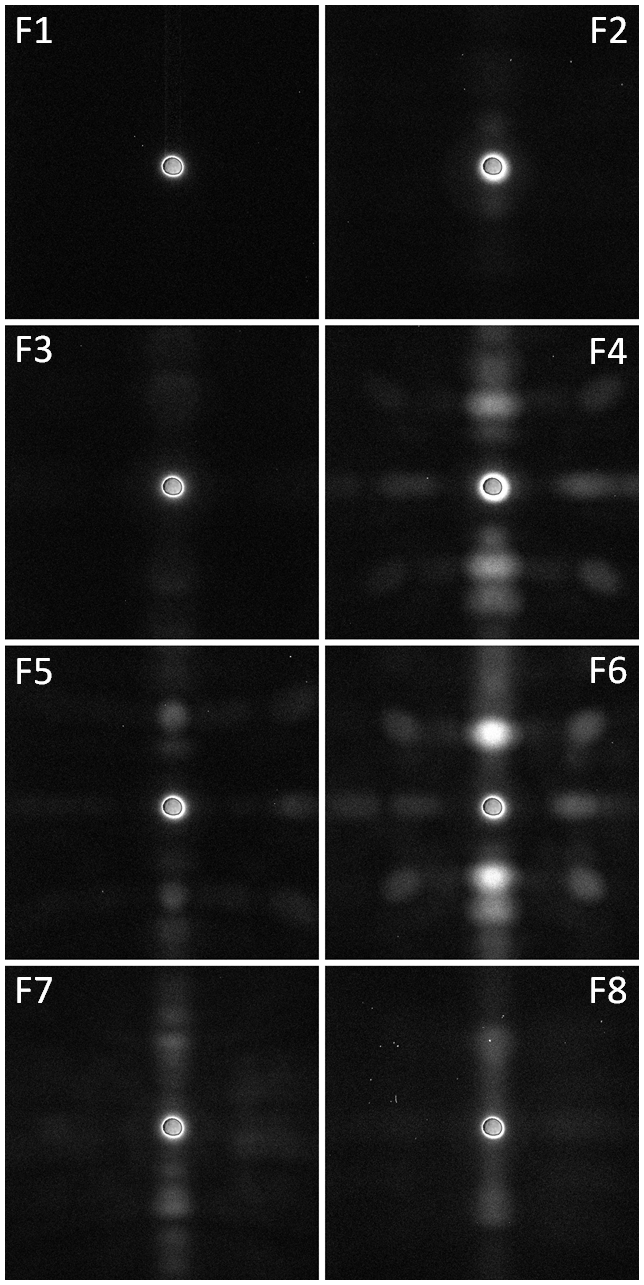}
\caption{Images of Vesta acquired during the {\it RC1} campaign show multiple ghost images distributed around the center in a diffraction pattern. The images have been enhanced in identical fashion to bring out the ghosts; black is zero intensity, white is 0.35\% of the average Vesta intensity. The correctly exposed F1 image of Vesta is superposed on the center of all images, showing the extent of stray light around the disk. Bright specks are cosmic rays, mostly visible in the F8 image because of the very long exposure time.}
\label{fig:ghosts}
\end{figure}

On approach to Vesta, FC2 exposures were acquired as part of the {\it Rotational Characterization \#1} ({\it RC1}) campaign in all filters with the asteroid in the center of the image. This provided a perfect opportunity to assess the effects of in-field stray light. This type of stray light had been identified in pre-launch calibration images \citep{S11}, but had not yet been demonstrated in-flight. The most important factor contributing to stray light is the closeness of the filters to the optical plane, which leads to multiple reflections between the interference-type narrow-band filters and the highly reflective CCD surface. The CCD is front-lit and acts as a grating due to the presence of anti-blooming and other gates. This particular type of stray light affects images acquired through the narrow-band filters but not images taken through the clear filter, as the latter is not of the interference type. Pre-launch images of a small pinhole observed through a collimator show multiple ghost images around the pinhole arranged in a diffraction pattern. During {\it RC1}, Vesta had approximately the same apparent size as the pinhole ($\sim$55~pixels). Around the Vesta disk we find the same type of ghost images (Fig.~\ref{fig:ghosts}). Scaling the {\it RC1} images brightness in identical fashion, we find them similar to the pinhole images in every aspect, both qualitatively and quantitatively (compare Fig.~42 in \citealt{S11}). Figure~\ref{fig:ghosts} shows subtle differences in shape and strength of the stray light pattern. Filters F4 and F6 have relatively strong interference patterns, while F2 and F4 have an extended halo around the central disk. The internal reflections of the standard stars and Saturn mostly fall outside the photometry apertures, and thus the radiometric calibration derived from these observations is only valid for images free of in-field stray light. As such, it can be expected that the reflectance reconstructed from Vesta narrow-band images calibrated with the coefficients provided in Table~\ref{tab:filters} is overestimated. A description of a method for correcting in-field stray light is deferred to a follow-up paper. We note that the flat fields as recorded before launch are also affected by in-field stray light, and cannot be used for calibration purposes. Their revision will be dealt with in the same paper.

\subsection{Out-of-field}
\label{sec:out-of-field}

\begin{figure}
\centering
\includegraphics[width=\textwidth,angle=0]{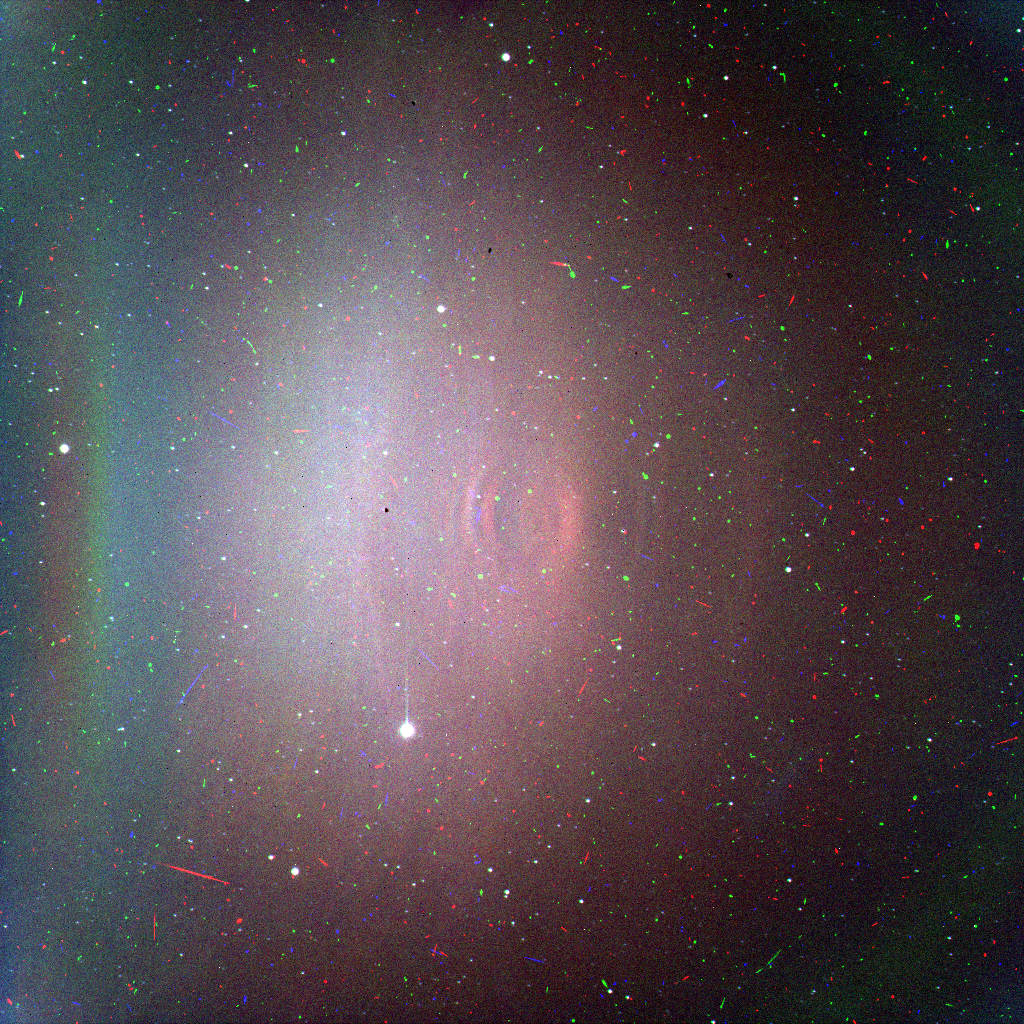}
\caption{Contrast-enhanced color image of out-of-field stray light with the Sun as a light source. The (R, G, B) color channels represent filters (F5, F6, F3). Images were not divided by a flat field. The numerous colored specks and streaks are cosmic rays; only white objects are stars.}
\label{fig:color_stray_light}
\end{figure}

To characterize and quantify the potential out-of-field stray light contribution to the image quality, FC2 observed the deep sky using the Sun as a stray light source (Sec~\ref{sec:stray_light_campaign}). The resulting images show a distinctive pattern of stray light that increases in strength with decreasing Sun angle. Stray light elevates the signal level over the full frame, slightly more so in the center. The color composite in Fig.~\ref{fig:color_stray_light} shows the stray light contribution at various wavelengths. The center of the stray light pattern is most pronounced in the near-IR (shown as red in the color image), because the coating on the inside of the baffle is more reflective in the near-IR ($>800$~nm) than in the visible wavelength range. Some aspects of the stray light pattern are wavelength dependent. For example, the vertical streak at left shifts leftward with increasing wavelength.

\begin{figure}
\centering
\includegraphics[width=8cm,angle=0]{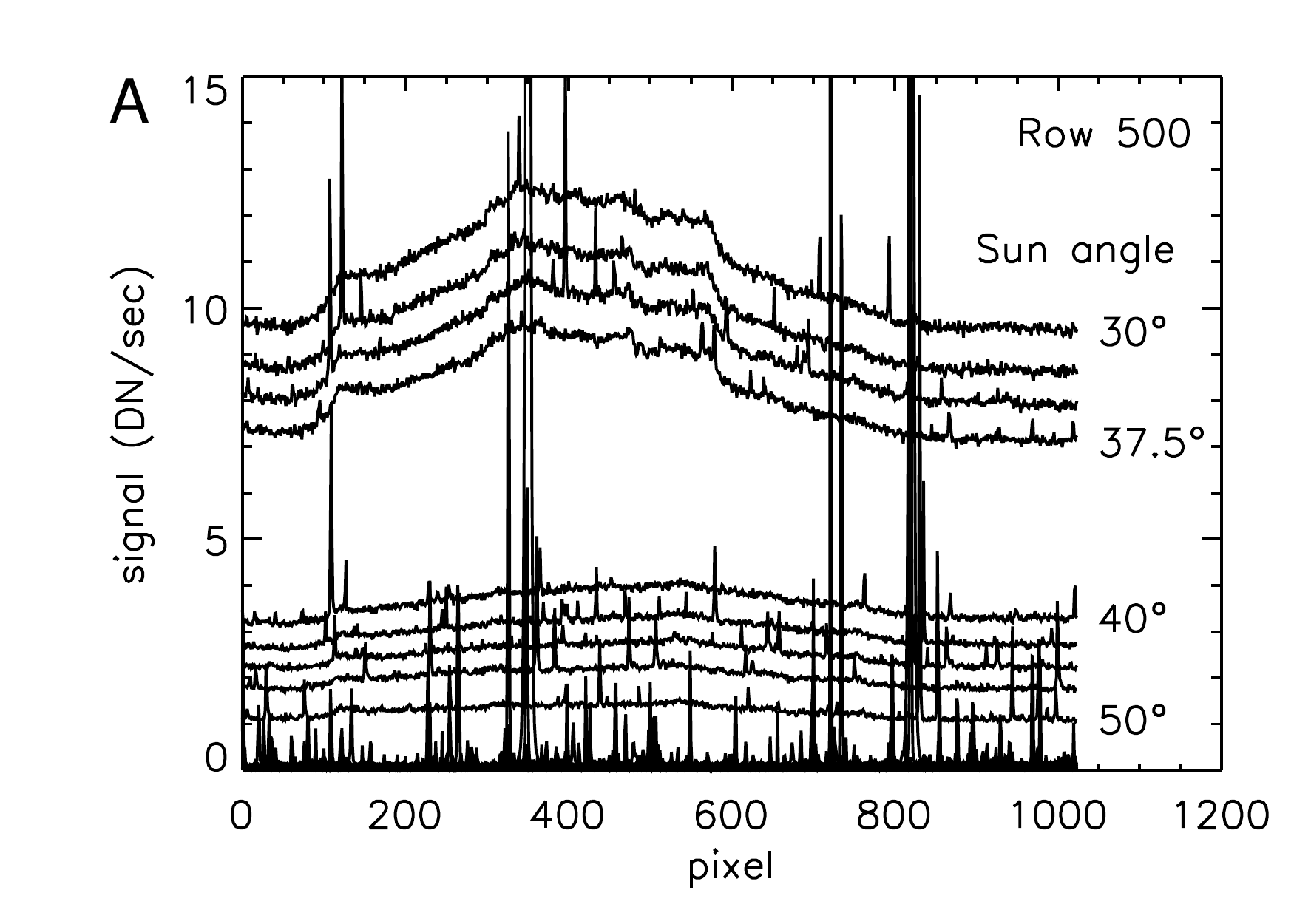}
\includegraphics[width=8cm,angle=0]{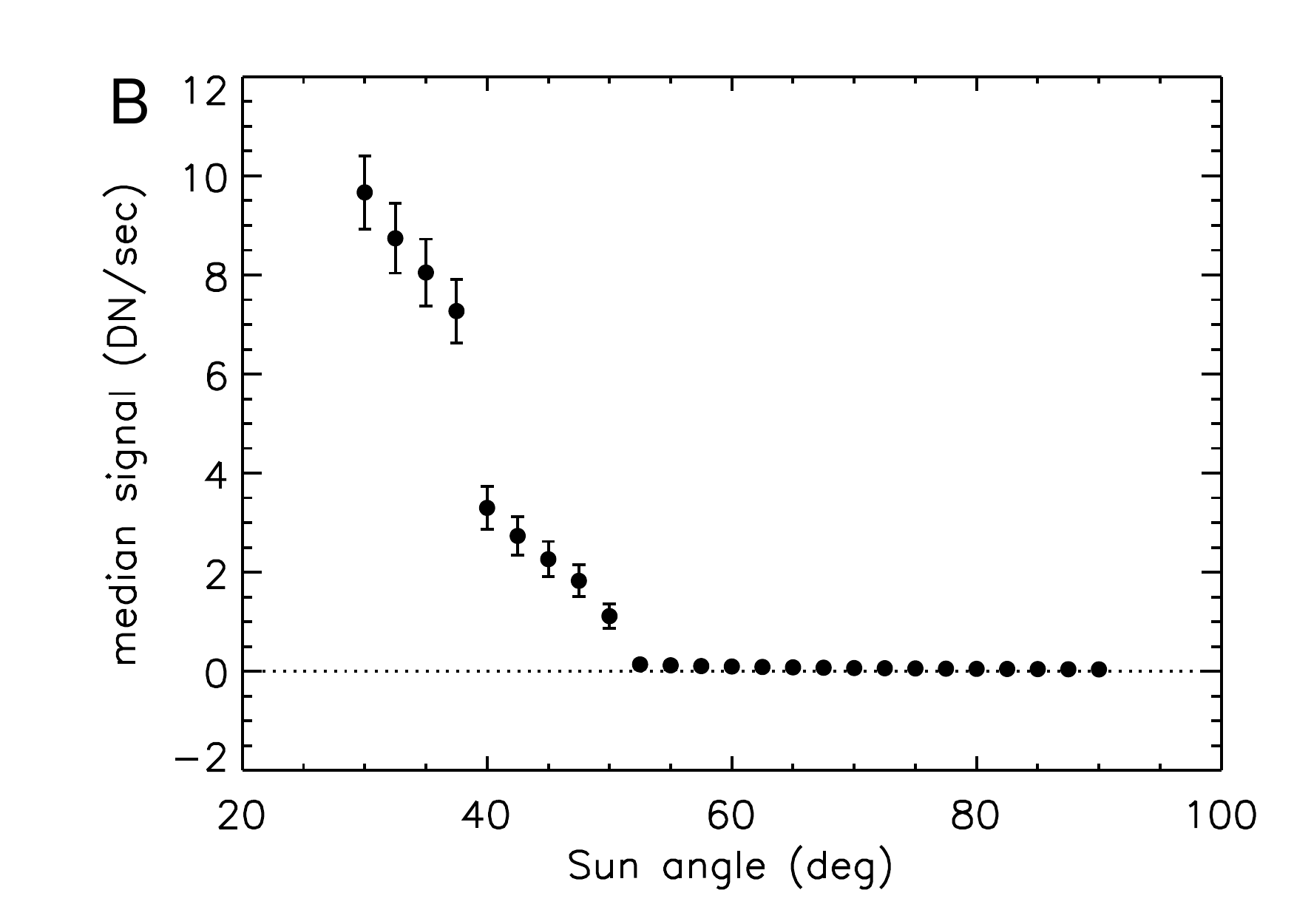}
\caption{Characterization of out-of-field stray light in 100~s clear filter exposures with the Sun as a light source. {\bf A}:~Stray light profiles (row 500) at different Sun angles. {\bf B}:~Stray light contribution (median of full frame) as a function of Sun angle. The error bars denote the photon noise level.}
\label{fig:F1_stray_light}
\end{figure}

While in-field stray light is a major nuisance for the color filters, the contribution of out-of-field stray light is relatively small. Figure~\ref{fig:F1_stray_light} shows that the stray light contribution in the clear filter stayed below 13~DN~s$^{-1}$ (Vesta flux at zero phase angle is around $10^6$~DN~s$^{-1}$). Above $50^\circ$ Sun angle stray light is virtually absent; below it the level increases steadily. There are two angles at which the stray light jumps to higher levels, the first between $52.5^\circ$ and $50^\circ$ and the second between $40^\circ$ and $37^\circ$. These "critical angles" must be associated with structures inside the baffle, like field stops. With the availability of this ``extended source'' we considered testing the quality of our flat fields, in particular the pixel-to-pixel variations, as pioneered by \citet{M99}, However, the stray light levels were too low for this purpose, the exposure times being too short, with the photon noise level exceeding these variations.

As mentioned above, when searching for moons, the asteroid itself can act as a source of stray light. A moon search around Vesta was performed on approach, and had the asteroid just outside of the field of view, at an angle much smaller than explored in this experiment ($3^\circ$ versus $30^\circ$). The levels of stray light experienced during the moon search were indeed much higher than described here, but are reported elsewhere (e.g. \citealt{McF11}). Note that also for regular exposures of Vesta's surface, the illuminated surface outside of the FOV will act as a source of stray light. While we were not able to quantify it in the lab, this contribution is probably small.

\section{Conclusions}
\label{sec:discussion}

We present a new calibration for images of both Dawn Framing Cameras (FC1 and FC2) on basis of an analysis of in-flight data, acquired {\it en route} to Vesta. The proposed radiometric and geometric calibration represents an improvement over the ground calibration as reported by \citet{S11}. The absolute radiometric calibration of point sources is accurate to a few percent, despite the reduced pixel fill factor of the FC CCD. Geometric (radial) distortion is very small, and could be determined to an accuracy of a little more than a tenth of a pixel in the image corners. Dark current is monitored on a regular basis, and remains very low at the camera operational temperatures. Residual charge, present on the CCD at the start of an exposure was confirmed for FC1, but not detected for FC2. The performance of both cameras is very similar, except for their geometric distortion and residual charge characteristics.

We emphasize that the radiometric calibration is only valid for point sources, and not extended sources, like Vesta. The FC suffers from significant in-field stray light when observing extended objects. This was already expected on basis of laboratory images \citep{S11}, and the {\it RC1} images confirm that all narrow-band images of Vesta are affected. We estimate that in-field stray light may contribute up to 15\% of the signal, depending on the filter, when Vesta fills the field-of-view. The actual stray light pattern depends on the scene observed, but is generally concentrated in the center of the frame. When constructing color images from narrow-band images, this may result in spurious color gradients across the image. The FC stray light patterns are similar to those affecting the Messenger spacecraft camera \citep{D11}, but, in contrast, do not show a clear increase of strength with wavelength. Validating our calibration with Vesta data requires a stray light correction, as all narrow-band images of Vesta are understood to be affected. Such a correction is beyond the scope of this paper, and will be addressed in one or more separate papers. Tightly connected to the in-field stray light problem is the matter of flat fielding. Flat fields constructed from images of the inside of an integrating sphere are affected by the same type of in-field stray light as the Vesta images, giving them a distinct ``curved'' appearance \citep{S11}. Using these non-flat fields in the calibration process will lead to incorrect results, as stray light needs to be subtracted rather than divided out. The discussion on which flat fields to use is closely tied to the stray light problem, and must also be deferred.

\section*{Acknowledgements}

The authors thank S. Besse and an anonymous referee for their excellent comments that helped to improve the manuscript. This research has made use of the SIMBAD database\footnote{\url{http://simbad.u-strasbg.fr/simbad/}}, operated at CDS, Strasbourg, France.



\bibliography{calib}

\newpage
\clearpage

\begin{table}
\centering
\caption{Targeted observations during the {\it ICO} campaign ($n$ is number of images acquired).}
\vspace{5mm}
\begin{tabular}{llll}
\hline
Model & Target & $n$ & Comments \\
\hline
FC1	& 20 Cep	 & 20  & 4 images intentionally highly overexposed \\
FC1	& Vega       & 96  & Vega outside the FOV in 12 out of 96 images \\
FC1	& 73 Cet   & 16  & \\
FC1	& 51 Peg  & 16  & \\
FC1	& NGC 3532   & 104 & $2\times2$ mosaic plus center frame, \\
    &            &     & 2 images intentionally highly overexposed \\
FC2	& 20 Cep  & 20  & 4 images intentionally highly overexposed \\
FC2	& Vega       & 96  & \\
FC2	& 42 Peg  & 16  & \\
FC2	& 51 Peg  & 16  & \\
FC2	& NGC 3532   & 104 & $2\times2$ mosaic plus center frame, \\
    &            &     & 2 images intentionally highly overexposed \\
FC2	& Cassiopeia & 96  & $1\times3$ mosaic \\
FC2	& Saturn     & 64  & \\
\hline
\end{tabular}
\label{tab:ICO}
\end{table}


\begin{table}
\centering
\caption{Targets of the pointed semi-annual checkout ($n$ is number of images acquired). Mission target is valid for the Mars-Vesta cruise phase; for the Vesta-Ceres cruise phase the mission target is Ceres, with no backup target.}
\vspace{5mm}
\begin{tabular}{lllll}
\hline
Target type    & Prime target & Backup target & $n$ & Comments \\
\hline
Standard star  & Vega         & Alkaid ($\eta$ UMa) & 85 & 81 windowed, 4 full frame \\
               &              &                     &    & intentionally overexposed \\
Solar analog   & HR 2290      & 18 Sco              & 20 & windowed \\
Mission target & 4 Vesta      & 1 Ceres             & 16 & windowed \\
Star field     & 20 Cep       & NGC 3532            & 20 & full frame \\
\hline
\end{tabular}
\label{tab:semi-annual}
\end{table}


\begin{table}
\centering
\caption{Radiometric calibration parameters: Filter characteristics, absolute responsivity, and effective solar flux for both cameras and all filters. $\lambda_{\rm cen}$ and $\lambda_{\rm eff}$ are the filter band center and effective wavelength, respectively, in nm. $\Delta\lambda$ is the FWHM of the transmission profile, the boundaries of which are those indicated for $\lambda_{\rm eff}$. $c$ is the calibration correction factor. The radiance is obtained by dividing the pixel signal (in DN s$^{-1}$) by $\mathcal{R}$. The result of this division has units of W m$^{-2}$ nm$^{-1}$ sr$^{-1}$, except for F1, for which it is W m$^{-2}$ sr$^{-1}$. $F_\odot$ is the filter-specific effective solar flux in W m$^{-2}$ nm$^{-1}$ at 1~AU. The responsivity is camera-specific only for F8.}
\vspace{5mm}
\begin{tabular}{llllllll}
\hline
Filter & Model & $\lambda_{\rm cen}$ & $\Delta\lambda$ & $\lambda_{\rm eff}$ & $F_\odot$ & $c$ & $\mathcal{R}$ \\
\hline
F1 &     & 735 &     & $732$             &       & 1.11 & $5.12\times10^4$ \\
F2 &     & 549 & 44  & $555^{+15}_{-28}$ & 1.863 & 1.13 & $1.93\times10^6$ \\
F3 &     & 749 & 45  & $749^{+22}_{-22}$ & 1.274 & 1.10 & $3.85\times10^6$ \\
F4 &     & 919 & 46  & $917^{+24}_{-21}$ & 0.865 & 1.10 & $1.82\times10^6$ \\
F5 &     & 978 & 87  & $965^{+56}_{-29}$ & 0.785 & 1.10 & $1.76\times10^6$ \\
F6 &     & 829 & 37  & $829^{+18}_{-18}$ & 1.058 & 1.15 & $2.47\times10^6$ \\
F7 &     & 650 & 43  & $653^{+18}_{-24}$ & 1.572 & 1.13 & $3.22\times10^6$ \\
F8 & FC1 & 428 & 41  & $438^{+10}_{-30}$ & 1.743 & 1.30 & $1.95\times10^5$ \\
   & FC2 & 428 & 41  & $438^{+10}_{-30}$ & 1.743 & 1.45 & $2.18\times10^5$ \\
\hline
\end{tabular}
\label{tab:filters}
\end{table}


\begin{table}
\centering
\caption{Geometric distortion characteristics of each filter (focal length $f$, radial distortion parameter $k_1$, and the IFOV), found by fitting $\sim$100 star positions in {\it ICO Performance} (FC1 and FC2) and {\it DC041} (FC2 only) images of the field around 20~Cep ($p_1 = p_2 = 0$). Two images were acquired per filter, so $n = 2$ for FC1 and $n = 4$ for FC2. The numbers in brackets in the FC2 $k_1$ column are predictions based on a linear fit to the measurements (Fig.~\ref{fig:geom_dist}). The FC1 $k_1$ is assumed to be constant for all color filter except F8. The focal length values are based on a pixel size in the $x$-direction of 14.004~$\mu$m. IFOV is given in radians for the $x$- and $y$-direction.}
\vspace{5mm}
\begin{tabular}{lllll}
\hline
Model & Filter & $f$ & $k_1$ & IFOV \\
 &  & (mm) & ($10^{-6}$~mm$^{-2}$) & ($10^{-5}$) \\
\hline
FC1 & F1 & $150.074\pm0.004$ & $7.6\pm0.1$ & $9.3242\times9.3184$ \\
    & F2 & $150.081\pm0.001$ & $7.4\pm0.6$ (7.7) & $9.3238\times9.3179$ \\
    & F3 & $150.047\pm0.002$ & $7.7\pm0.2$ (7.7) & $9.3259\times9.3200$ \\
    & F4 & $150.130\pm0.007$ & $7.4\pm0.9$ (7.7) & $9.3208\times9.3149$ \\
    & F5 & $150.164\pm0.004$ & $7.4\pm0.3$ (7.7) & $9.3187\times9.3128$ \\
    & F6 & $150.073\pm0.020$ & $8.0\pm1.3$ (7.7) & $9.3243\times9.3184$ \\
    & F7 & $150.040\pm0.011$ & $7.3\pm1.3$ (7.7) & $9.3264\times9.3205$ \\
    & F8 & $150.394\pm0.010$ & $1.5\pm1.0$ & $9.3044\times9.2985$ \\
FC2 & F1 & $150.074\pm0.006$ & $8.4\pm0.4$ & $9.3242\times9.3184$ \\
    & F2 & $150.105\pm0.012$ & $6.8\pm1.2$ (6.7) & $9.3223\times9.3164$ \\
    & F3 & $150.044\pm0.005$ & $9.0\pm0.4$ (8.4) & $9.3261\times9.3202$ \\
    & F4 & $150.119\pm0.008$ & $9.8\pm0.6$ (10.0) & $9.3215\times9.3156$ \\
    & F5 & $150.158\pm0.011$ & $9.4\pm1.3$ (10.3) & $9.3190\times9.3131$ \\
    & F6 & $150.081\pm0.013$ & $8.6\pm1.3$ (9.2) & $9.3238\times9.3179$ \\
    & F7 & $150.055\pm0.002$ & $7.2\pm0.4$ (7.6) & $9.3254\times9.3195$ \\
    & F8 & $150.380\pm0.010$ & $5.3\pm1.7$ (5.6) & $9.3053\times9.2994$ \\
\hline
\end{tabular}
\label{tab:geom_dist}
\end{table}

\newpage
\clearpage

\end{document}